\documentclass[11pt]{article}

\usepackage[a4paper,text={16cm,25.7cm},centering]{geometry}
\usepackage{bm,amssymb,amsmath,amsfonts,graphicx,xcolor}
\usepackage[numbers, comma, sort&compress]{natbib}
\usepackage{hyperref}
\usepackage[utf8]{inputenc}
\usepackage{epstopdf}

\newcommand{\beginsupplement}{%
        \setcounter{table}{0}
        \renewcommand{\thetable}{S\arabic{table}}%
        \setcounter{figure}{0}
        \renewcommand{\thefigure}{S\arabic{figure}}%
        \setcounter{equation}{0}
        \renewcommand{\theequation}{S\arabic{equation}}
     }

\begin{document}

\begin{flushleft}
{\Large
\textbf\newline{Quantifying the impact of a periodic presence of antimicrobial on resistance evolution in a homogeneous microbial population of fixed size} 
}
\newline
\\
Loïc Marrec\textsuperscript{1},
Anne-Florence Bitbol\textsuperscript{1*}
\\
\bigskip
\textbf{1} Sorbonne Université, CNRS, Laboratoire Jean Perrin (UMR 8237), F-75005 Paris, France
\\
\bigskip

* anne-florence.bitbol@sorbonne-universite.fr

\end{flushleft}
\section*{Abstract}

The evolution of antimicrobial resistance generally occurs in an environment where antimicrobial concentration is variable, which has dramatic consequences on the microorganisms' fitness landscape, and thus on the evolution of resistance. 

We investigate the effect of these time-varying patterns of selection within a stochastic model. We consider a homogeneous microbial population of fixed size subjected to periodic alternations of phases of absence and presence of an antimicrobial that stops growth. Combining analytical approaches and stochastic simulations, we quantify how the time necessary for fit resistant bacteria to take over the microbial population depends on the alternation period. We demonstrate that fast alternations strongly accelerate the evolution of resistance, reaching a plateau for sufficiently small periods. Furthermore, this acceleration is stronger in larger populations. For asymmetric alternations, featuring a different duration of the phases with and without antimicrobial, we shed light on the existence of a minimum for the time taken by the population to fully evolve resistance. The corresponding dramatic acceleration of the evolution of antimicrobial resistance likely occurs in realistic situations, and may have an important impact both in clinical and experimental situations. 

\section*{Introduction}
The discovery of antibiotics and antivirals has constituted one of the greatest medical advances of the twentieth century, allowing many major infectious diseases to be treated. However, with the increasing use of antimicrobials, pathogenic microorganisms tend to become resistant to these drugs. Antimicrobial resistance has become a major and urgent problem of public health worldwide~\cite{WHO,AMR}.

Mutations that confer antimicrobial resistance are often associated with a fitness cost, i.e. a slower reproduction~\cite{Borman96,Andersson10,zurWiesch11}. Indeed, the acquisition of resistance generally involves either a modification of the molecular target of the antimicrobial, which often alters its biological function, or the production of specific proteins, which entails a metabolic cost~\cite{Andersson10}. However, resistant microorganisms frequently acquire subsequent mutations that compensate for the initial cost of resistance. These microorganisms are called ``resistant-compensated''~\cite{Schrag97,Levin00,Paulander07,deSousa15}. The acquisition of resistance is therefore often irreversible, even if the antimicrobial is removed from the environment~\cite{Schrag97,Andersson10}. 

In the absence of antimicrobial, the adaptive landscape of the microorganism, which represents its fitness (i.e. its reproduction rate) as a function of its genotype, involves a valley, since the first resistance mutation decreases fitness, while compensatory mutations increase it. However, this fitness valley, which exists in the absence of antimicrobial, disappears above a certain concentration of antimicrobial, as the growth of the antimicrobial-sensitive microorganism is impaired. Thus, the adaptive landscape of the microorganism depends drastically on whether the antimicrobial is present or absent. Taking into account this type of interaction between genotype and environment constitutes a fundamental problem, even though most experiments have traditionally focused on comparing different mutants in a unique environment~\cite{Taute14}. In particular, recent theoretical analyses show that variable adaptive landscapes can have a dramatic evolutionary impact~\cite{Mustonen08,Rivoire11,Melbinger15,Desponds16,Wienand17}.

How do the timescales of evolution and variation in the adaptive landscape compare and interact? What is the impact of the time variability of the adaptive landscape on the evolution of antimicrobial resistance? 
In order to answer these questions, we construct a minimal model retaining the fundamental aspects of antimicrobial resistance evolution. Focusing on the case of a homogeneous microbial population of fixed size, we perform a complete stochastic study of \textit{de novo} resistance acquisition in the presence of periodic alternations of phases of absence and presence of an antimicrobial that stops growth. These alternations can represent, for example, a treatment where the concentration within the patient falls under the Minimum Inhibitory Concentration (MIC) between drug intakes~\cite{Regoes04}. Combining analytical and numerical approaches, we show that these alternations substantially accelerate the evolution of resistance with respect to the cases of continuous absence or continuous presence of antimicrobial, especially for larger populations. We fully quantify this effect and shed light on the different regimes at play. For asymmetric alternations, featuring a different duration of the phases with and without antimicrobial, we demonstrate the existence of a minimum for the time taken by the population to fully evolve resistance, occurring when both phases have durations of the same order. This realistic situation dramatically accelerates the evolution of resistance. Finally, we discuss the implications of our findings, in particular regarding antimicrobial dosage.

\section*{Model}

\begin{figure}[h t b]
\centering
\includegraphics[width=\textwidth]{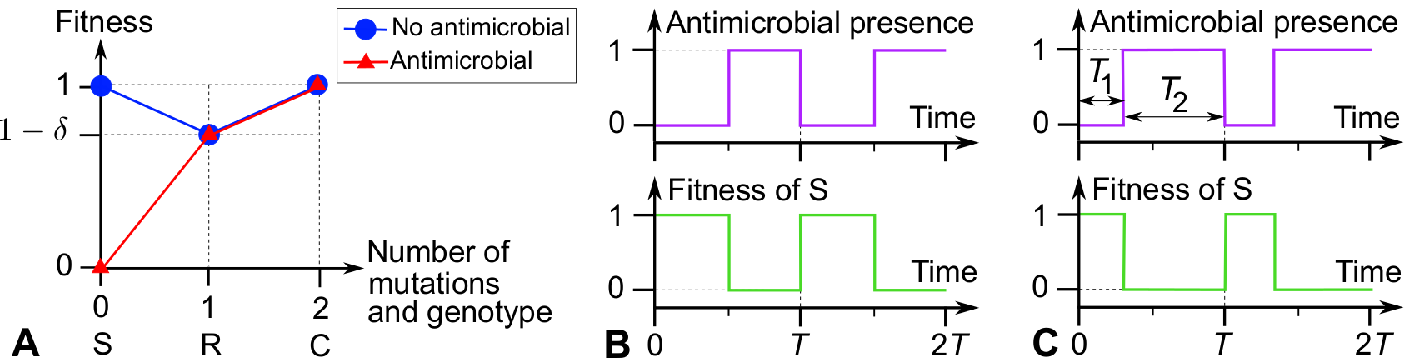}\\
\vspace{0.2cm}
\caption{\label{Fig1}\textbf{Model.} (A) Adaptive landscapes in the presence and in the absence of antimicrobial. Genotypes are indicated by the number of mutations from the sensitive microorganism, and by initials: S: sensitive; R: resistant; C: resistant-compensated. \protect\\(B) and (C) Periodic presence of antimicrobial, and impact on the fitness of S (sensitive) microorganisms: (B) Symmetric alternations; (C) Asymmetric alternations.}
\end{figure}

The action of an antimicrobial drug can be quantified by its MIC, the minimum concentration that stops the growth of a microbial population~\cite{Andersson10}. We focus on biostatic antimicrobials, which stop microbial growth (vs. biocidal antimicrobials, which kill microorganisms). We model the action of the antimicrobial in a binary way: below the MIC (``absence of antimicrobial''), growth is not affected, while above it (``presence of antimicrobial''), sensitive microorganisms cannot grow at all. The usual steepness  of pharmacodynamic curves around the MIC~\cite{Regoes04} justifies our simple binary approximation, and we also present an analysis of the robustness of this hypothesis (Supplementary Material, Section~\ref{SI_Robust}). Within this binary approximation, there are two adaptive landscapes. Assuming that the drug fully stops the growth of sensitive microorganisms, but does not affect that of resistant ones, and considering compensatory mutations that fully restore fitness, these two adaptive landscapes can be described by a single parameter $\delta$, representing the fitness cost of resistance (Fig.~\ref{Fig1}A). We focus on asexual microorganisms, and fitness simply denotes the division rate of these organisms. The fitness of sensitive microorganisms in the absence of antimicrobials is taken as reference. In this framework, we investigate the impact of a periodic presence of antimicrobial, assuming that the process starts without antimicrobial (Fig.~\ref{Fig1}B-C). 

We denote by $\mu_1$ and $\mu_2$ the mutation rates (or mutation probabilities upon each division) for the mutation from S to R and for the one from R to C, respectively. In several actual situations, the effective mutation rate towards compensation tends to be higher than the one towards the return to sensitivity, since multiple mutations can compensate for the initial cost of resistance~\cite{Levin00,Paulander07,Hughes15}. Therefore, we do not take into account back-mutations. Still because of the abundance of possible compensatory mutations,  generally  $\mu_1\ll\mu_2$~\cite{Levin00,Poon05}. We present general analytical results as a function of $\mu_1$ and $\mu_2$, and analyze in more detail the limit $\mu_1\ll\mu_2$, especially in simulations. All notations introduced are summed up in Table~\ref{Not_table}.

We focus on a homogeneous microbial population of fixed size $N$, which can thus be described in the framework of the Moran process~\cite{Moran58,Ewens79}, where fitnesses are relative (see Supplementary Material, Section~\ref{SI_Moran} and Fig.~\ref{MP}). Assuming a constant size simplifies the analytical treatment and is appropriate for instance to describe turbidostat experiments, where the dilution rate is adjusted so that turbidity (and hence population size) is constant~\cite{Myers44}. If a population only features sensitive individuals (with zero fitness) in the presence of antimicrobial, we consider that no division occurs, and the population remains static. We always express time in number of generations, which corresponds (unless no cell can divide) to the number of Moran steps divided by the population size $N$. 

Throughout, we start from a microbial population where all individuals are S (sensitive), and we focus on the time $t_\mathrm{C}^\mathrm{f}$ it takes for the C (resistant-compensated) type to fix in the population, i.e. to take over the population. Then, the population has fully evolved resistance \textit{de novo}. 

\section*{Results}

\subsection*{A periodic presence of antimicrobial can drive resistance evolution}

In this section, we study how alternations of absence and presence of antimicrobial can drive the \textit{de novo} evolution of resistance. We present analytical predictions for the time needed for the population to evolve resistance, and then we compare them to numerical simulation results.

We first focus on the rare mutation regime $N\mu_1\ll 1$, where at most one mutant lineage exists in the population at each given time. The frequent mutation regime is briefly discussed, and more detail regarding the appropriate deterministic treatment in this regime is given in Supplementary Material, Section~\ref{SI_Det}. Here, we consider the case of symmetric alternations with period $T$ (Fig.~\ref{Fig1}B). Asymmetric alternations (Fig.~\ref{Fig1}C) will be discussed later.

\subsubsection*{Time needed for resistant microorganisms to start growing}

Resistant (R) mutants can only appear during phases without antimicrobial. Indeed, mutations occur upon division, and sensitive (S) bacteria cannot divide in the presence of antimicrobial (Fig.~\ref{Fig1}). However, R mutants are less fit than S individuals without antimicrobial. Hence, the lineage of an R mutant will very likely disappear, unless it survives until the next addition of antimicrobial. More precisely, without antimicrobial, the fixation probability $p_\mathrm{SR}$ of a single R mutant with fitness $1-\delta$, in a population of size $N$ where all other individuals are of type S and have fitness 1, is $\sim\!1/N$ if the mutation from S to R is effectively neutral ($N\delta\ll 1$), and $\sim\!\delta e^{-N\delta}$ if $\delta\ll 1$ and $N\delta\gg 1$~\cite{Ewens79}. Let us denote by $\tau_\mathrm{R}^\mathrm{d}$ the average time an R lineage would drift before going extinct without antimicrobial~\cite{Ewens79} (see Supplementary Material, Section~\ref{SI_Moran}). If antimicrobial is added while R mutants exist in the population, i.e. within $\sim\!\tau_\mathrm{R}^\mathrm{d}$ after a mutation event, then the R population will grow fast and fix, since S individuals cannot divide with antimicrobial. Hence, each time antimicrobial is added, any R lineage that was destined for extinction without antimicrobial  but that survived until the addition of drug  is rescued. Through this phenomenon, periodic alternations of absence and presence of antimicrobial can substantially accelerate resistance evolution: we will quantify this effect. Note that here, we disregard the very few R lineages destined for fixation without antimicrobial, because we aim to study the acceleration of resistance evolution due to the alternations. The spontaneous evolution of resistance without antimicrobial is discussed and compared to our alternation-driven process in the Supplementary Material, Section~\ref{SI_Valley}.

\begin{figure}[h!]
\centering
  \includegraphics[width=1.0\linewidth]{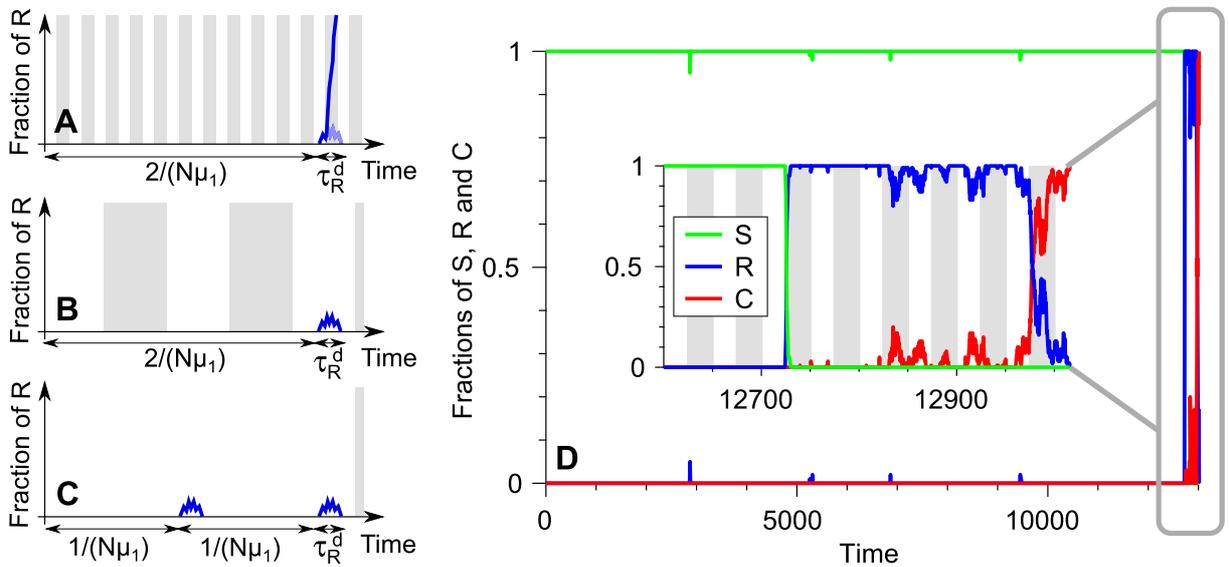}
  \quad
\caption{\textbf{Alternation-driven evolution of antimicrobial resistance.} (A-C) Sketches illustrating the three different regimes for the half-period $T/2$ of the alternations of antimicrobial absence (white) and presence (gray). The fraction of resistant (R) microorganisms in the population is plotted versus time (blue curves). R mutants can only appear without antimicrobial. (A) $T/2 \ll \tau_\mathrm{R}^\mathrm{d}$, where $\tau_\mathrm{R}^\mathrm{d}$ is the average extinction time of the lineage of an R mutant without antimicrobial. The first R lineage that appears is expected to live until the next addition of antimicrobial and is then rescued. (B) $\tau_\mathrm{R}^\mathrm{d} \ll T/2\ll 1/(N\mu_1)$, where $1/(N\mu_1)$ is the average time between the appearance of two independent R mutants without antimicrobial. (C) $T/2 \gg 1/(N\mu_1)$. In (B) and (C), not all R lineages live until the next addition of antimicrobial, and in (C) multiple R lineages arise within a half-period. (D) Example of a simulation run. The fractions of S, R and C microorganisms are plotted versus time. Inset: end of the process, with full resistance evolution. As in (A-C), antimicrobial is present during the gray-shaded time intervals (shown only in the inset given their duration). Parameters: $\mu_1=10^{-5}$, $\mu_2=10^{-3}$, $\delta=0.1$, $N=10^2$ and $T=50$ (belonging to regime B).}
  \label{regimes_and_example}
\end{figure}

It is crucial to calculate the average waiting time $t^\mathrm{a}_\mathrm{R}$ until an R lineage is rescued by the addition of antimicrobial. Indeed, this constitutes the key step of alternation-driven resistance takeover. Three timescales impact $t^\mathrm{a}_\mathrm{R}$. The first one is the timescale of the environment, namely the half-period $T/2$. The two other ones are intrinsic timescales of the evolution of the population without antimicrobial: the average time between the appearance of two independent R mutants, $1/(N\mu_1)$, and the average lifetime $\tau_\mathrm{R}^\mathrm{d}$ of the lineage of an R mutant destined for extinction without antimicrobial. Note that $\tau_\mathrm{R}^\mathrm{d}$ is  generally  quite short. Indeed, $\tau_\mathrm{R}^\mathrm{d}\approx\log N$ for large $N$ if $\delta=0$, and $\tau_\mathrm{R}^\mathrm{d}$ decreases as $\delta$ increases, as deleterious R mutants are out-competed by S microorganisms; for instance, $\tau_\mathrm{R}^\mathrm{d}\approx 2.6$ generations if $\delta=0.1$ in the limit where $N\gg 1$ and $N\delta \gg 1$~\cite{Ewens79} (see Supplementary Information, Section~\ref{SI_Moran}). Hence, in the rare mutation regime, $\tau_\mathrm{R}^\mathrm{d} \ll 1/(N\mu_1)$. What matters is how the environment timescale $T/2$ compares to these two evolution timescales (see Fig.~\ref{regimes_and_example}A-C). Our arguments based on comparing average timescales are approximate, but they yield explicit analytical predictions in each regime where timescales are separated, which we then test through numerical simulations.

(A) If $T/2 \ll \tau_\mathrm{R}^\mathrm{d}$ (Fig.~\ref{regimes_and_example}A): The lineage of the first R mutant that appears is likely to still exist upon the next addition of antimicrobial, and to be rescued, which yields $t^\mathrm{a}_\mathrm{R}=2/(N \mu_1)$. Indeed, mutations from S to R can only occur without antimicrobial, i.e. half of the time.  

(B) If $\tau_\mathrm{R}^\mathrm{d} \ll T/2\ll 1/(N\mu_1)$ (Fig.~\ref{regimes_and_example}B): At most one mutation yielding an R individual is expected within each half-period. The lineage of this mutant is likely to survive until the next addition of antimicrobial only if the mutant appeared within the last $\sim\tau_\mathrm{R}^\mathrm{d}$ preceding it, which has a probability $p=2\tau_\mathrm{R}^\mathrm{d}/T$. Hence, $t^\mathrm{a}_\mathrm{R}=2/(N \mu_1 p)=T/(N \mu_1 \tau_\mathrm{R}^\mathrm{d})$.

(C) If $T/2 \gg 1/(N\mu_1)$ (Fig.~\ref{regimes_and_example}C): Since the half-period is much larger than the time $1/(N\mu_1)$ between the appearance of two independent mutants without antimicrobial, several appearances and extinctions of R lineages are expected within one half-period. Hence, the probability that a lineage of R exists upon a given addition of antimicrobial is $q=N\mu_1\tau_\mathrm{R}^\mathrm{d}$,  which corresponds to the fraction of time during which R mutants are present in the phases without antimicrobial. Specifically, $q$ is the ratio of the average lifetime of the lineage of an R mutant destined for extinction without antimicrobial to the average time between the appearance of two independent R mutants without antimicrobial. Since additions of antimicrobial occur every $T$, we have $t^\mathrm{a}_\mathrm{R}= T/q=T/(N \mu_1 \tau_\mathrm{R}^\mathrm{d})$, which is the same as in case (B). In fact, the demonstration presented for case (C) also holds for case (B).

In conclusion, we obtain
\begin{equation}
t^\mathrm{a}_\mathrm{R}=\frac{T}{N \mu_1\min\left(\tau_\mathrm{R}^\mathrm{d},T/2\right)}.
\label{tR}
\end{equation}
Hence, if $T/2 \ll \tau_\mathrm{R}^\mathrm{d}$, $t^\mathrm{a}_\mathrm{R}$ is independent from the period $T$ of alternations, while if $T/2 \gg\tau_\mathrm{R}^\mathrm{d}$, $t^\mathrm{a}_\mathrm{R}$ is proportional to $T$.

\subsubsection*{Time needed for the population to fully evolve resistance}

We are interested in the average time $t_\mathrm{C}^\mathrm{f}$ it takes for the population to fully evolve resistance, i.e. for the C (resistant-compensated) type to fix. An example of the process is shown in Fig.~\ref{regimes_and_example}D. It takes on average $t^\mathrm{a}_\mathrm{R}$ for R mutants to be rescued by the addition of antimicrobial. Then they rapidly grow, since S individuals cannot divide. If the phase with antimicrobial is long enough, R mutants take over during this phase, with a probability 1 and an average fixation timescale $\tau^\mathrm{f}_\mathrm{R}\approx\log N$ for $N\gg 1$~\cite{Ewens79} (see Supplementary Material, Section~\ref{SI_Moran}). If $T/2\ll\tau^\mathrm{f}_\mathrm{R}$, fixation cannot occur within a single half-period, and the R lineage will drift longer, but its extinction remains very unlikely. Indeed,  while R individuals are the only ones that can divide with antimicrobial, we assume that they experience only a minor disadvantage without antimicrobial  ($1-\delta$ vs. 1, generally with $\delta\ll 1$~\cite{Andersson10}, see Fig.~\ref{Fig1}A). Hence, if $T/2\ll\tau^\mathrm{f}_\mathrm{R}$, and neglecting changes in frequencies in the absence of antimicrobial, R mutants will take $\sim\! 2 \tau^\mathrm{f}_\mathrm{R}$ to fix. 

Once the R type has fixed in the population, the appearance and eventual fixation of C mutants are independent from the presence of antimicrobial, since only S microorganisms are affected by it (see Fig.~\ref{Fig1}A). The first C mutant whose lineage will fix takes an average time $t_\mathrm{C}^\mathrm{a}=1/(N\mu_2\,p_\mathrm{RC})$ to appear once R has fixed, where $p_\mathrm{RC}$ is the fixation probability of a single C mutant in a population of size $N$ where all other individuals are of type R. In particular, if $N\delta\ll 1$ then $p_\mathrm{RC}=1/N$, and if $\delta\ll 1$ and $N\delta\gg 1$ then $p_\mathrm{RC}\approx\delta$~\cite{Ewens79} (see Supplementary Material, Section~\ref{SI_Moran}). The final step is the fixation of this successful C mutant, which will take an average time $\tau^\mathrm{f}_\mathrm{C}$, of order $N$ in the effectively neutral regime $N\delta\ll 1$, and shorter for larger $\delta$ given the selective advantage of C over R~\cite{Ewens79} (see Supplementary Material, Section~\ref{SI_Moran}). Note that we have assumed for simplicity that the fixation of R occurs before the appearance of the first successful C mutant, which is true if $t_\mathrm{C}^\mathrm{a}\gg\tau^\mathrm{f}_\mathrm{R}$, i.e. $1/(N\mu_2\,p_\mathrm{RC})\gg\log N$. This condition is satisfied if the second mutation is sufficiently rare. Otherwise, our calculation will slightly overestimate the actual result. 

Combining the previous results yields
\begin{align}
t_\mathrm{C}^\mathrm{f}\approx t^\mathrm{a}_\mathrm{R}+\tau^\mathrm{f}_\mathrm{R}+t_\mathrm{C}^\mathrm{a}+\tau^\mathrm{f}_\mathrm{C}\,, \label{TotTime}
\end{align}
where $t^\mathrm{a}_\mathrm{R}$ is given by Eq.~\ref{tR}, while $t_\mathrm{C}^\mathrm{a}=1/(N\mu_2\,p_\mathrm{RC})$, and $\tau^\mathrm{f}_\mathrm{R}\approx\log N$ and $\tau^\mathrm{f}_\mathrm{C}\lesssim N$. In the rare mutation regime, the contribution of the two fixation times $\tau^\mathrm{f}_\mathrm{R}$ and $\tau^\mathrm{f}_\mathrm{C}$ will be negligible. If in addition $\mu_1\ll\mu_2$, which is realistic (cf. Methods), then $t^\mathrm{f}_\mathrm{C}$ will be dominated by $t^\mathrm{a}_\mathrm{R}$. If $\mu_1\approx\mu_2$, $t^\mathrm{f}_\mathrm{C}$ will be dominated by $t^\mathrm{a}_\mathrm{R}$ if $T>\max\left(2\tau_\mathrm{R}^\mathrm{d},\,\,\tau_\mathrm{R}^\mathrm{d}/p_\mathrm{RC}\right)$. Indeed, if $T<2\tau_\mathrm{R}^\mathrm{d}$, using Eq.~\ref{tR} shows that the condition $t^\mathrm{a}_\mathrm{R}>t^\mathrm{a}_\mathrm{C}$ is then equivalent to $p_\mathrm{RC}>1/2$, which cannot be satisfied for $\delta\ll 1$. Hence, $T>2\tau_\mathrm{R}^\mathrm{d}$ is necessary to have $t^\mathrm{a}_\mathrm{R}>t^\mathrm{a}_\mathrm{C}$. But if $T>2\tau_\mathrm{R}^\mathrm{d}$ and $\mu_1\approx\mu_2$, the condition $t^\mathrm{a}_\mathrm{R}>t^\mathrm{a}_\mathrm{C}$ is equivalent to $T>\tau_\mathrm{R}^\mathrm{d}/p_\mathrm{RC}$. Beyond the regime $T>\max\left(2\tau_\mathrm{R}^\mathrm{d},\,\,\tau_\mathrm{R}^\mathrm{d}/p_\mathrm{RC}\right)$, the contribution of $t^\mathrm{a}_\mathrm{C}$ to $t_\mathrm{C}^\mathrm{f}$ will be important.

\subsubsection*{Comparison of analytical predictions and simulation results}

\begin{figure}[h!]
\centering
  \includegraphics[width=1.0\textwidth]{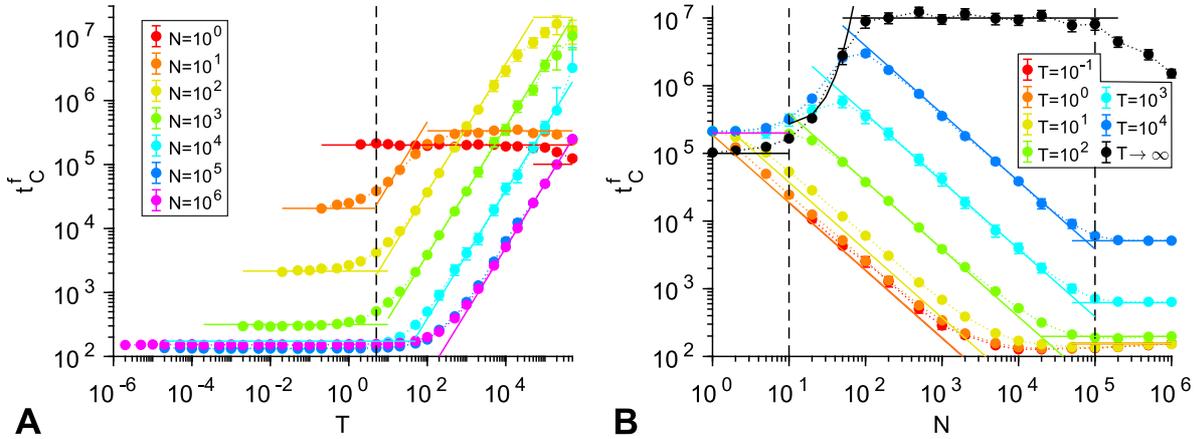}
  \quad
\caption{\textbf{Impact of symmetric alternations.} Fixation time $t_\mathrm{C}^\mathrm{f}$ of C (resistant-compensated) individuals in a population of $N$ individuals subjected to symmetric alternations of absence and presence of antimicrobial with period $T$. Data points correspond to the average of simulation results, and error bars (often smaller than markers) represent $95\%$ confidence intervals. 2 to $10^4$ replicate simulations were performed in each case (the smallest numbers of replicates were used for the largest populations, whose evolution is quasi-deterministic). In both panels, solid lines correspond to our analytical predictions in each regime. Parameter values: $\mu_1=10^{-5}$, $\mu_2=10^{-3}$, and $\delta=0.1$. (A) $t_\mathrm{C}^\mathrm{f}$ as function of $T$. Vertical dashed line: $T=2\tau_\mathrm{R}^\mathrm{d}$. (B) $t_\mathrm{C}^\mathrm{f}$ as function of $N$. Left vertical dashed line: limit of the neutral regime, $N=1/\delta$. Right vertical dashed line: limit of the deterministic regime, $N=1/\mu_1$. Horizontal purple line: analytical prediction for valley crossing by neutral tunneling in the presence of alternations (see Supplementary Material, Section~\ref{SI_Valley}). Black lines: analytical predictions for fitness valley crossing times in the absence of alternations (see Supplementary Material, Section~\ref{SI_Valley}).}
  \label{varyT_and_varyN}
\end{figure}

Fig.~\ref{varyT_and_varyN}A shows simulation results for the average total fixation time $t_\mathrm{C}^\mathrm{f}$ of C individuals in the population. This time is plotted as a function of the period $T$ of alternations for different population sizes $N$. As predicted above (see Eq.~\ref{tR}), we observe two regimes delimited by $T=2\tau_\mathrm{R}^\mathrm{d}$. If $T \ll 2\tau_\mathrm{R}^\mathrm{d}$, $t_\mathrm{C}^\mathrm{f}$ does not depend on $T$, while if $T \gg 2\tau_\mathrm{R}^\mathrm{d}$, it depends linearly on $T$. In Fig.~\ref{varyT_and_varyN}A, we also plot our analytical prediction from Eqs.~\ref{tR} and~\ref{TotTime} in these two regimes (solid lines). The agreement with our simulated data is excellent for small and intermediate values of $T$, without any adjustable parameter. Interestingly, the transition between these two regimes occurs for periods of about 5 generations, which would correspond to a few hours for typical bacteria, thus highlighting the practical importance of these two regimes. In Fig.~\ref{varyT_and_varyN}A, the smallest values reported for $t_\mathrm{C}^\mathrm{f}$ are of order 100 generations, corresponding to a few days, and are thus relevant to an actual treatment, while some other values are larger than the timescales involved in a treatment. Here, we quantitatively analyze the phenomena for a wide range of parameters. A more detailed comparison to actual situations, employing realistic values of population sizes and mutation rates, is presented in the Discussion. 

Importantly, Fig.~\ref{varyT_and_varyN}A shows that $t_\mathrm{C}^\mathrm{f}$ reaches a plateau for small $N$ and large $T$, which is not predicted by our analysis of the alternation-driven evolution of resistance. This plateau corresponds to the spontaneous fitness valley crossing process~\cite{Weissman09}, through which resistance  mutations appear and fix in the absence of drug. Note that such a plateau would also be reached for larger $N$, but for periods $T$ longer than those considered in Fig.~\ref{varyT_and_varyN}A (see Fig.~\ref{varyT_and_varyN}B, black lines). What ultimately matters is the shortest process among the alternation-driven one and the spontaneous valley-crossing one. In Fig.~\ref{varyT_and_varyN}A, horizontal solid lines at large $T$ represent our analytical predictions for the valley-crossing time (see Supplementary Material, Section~\ref{SI_Valley}). 

Fig.~\ref{varyT_and_varyN}B shows simulation results for $t_\mathrm{C}^\mathrm{f}$ as function of $N$ for different $T$. Again, solid lines represent our analytical predictions from Eqs.~\ref{tR} and~\ref{TotTime}, yielding excellent agreement for intermediate values of $N$, and for small ones at small $T$. In other regimes, resistance evolution is achieved by spontaneous valley crossing. In the limit $T\rightarrow\infty$ of continuous absence of antimicrobial (black data points in Fig.~\ref{varyT_and_varyN}B), only valley crossing can occur, and the black solid lines correspond to our analytical predictions for this process (see Supplementary Material, Section~\ref{SI_Valley}). 

Until now, we focused on the rare mutation regime. In the large-population, frequent-mutation regime $N\gg 1/\mu_1\gg 1$, the dynamics of the population can be well-approximated by a deterministic model with replicator-mutator differential equations~\cite{Traulsen05,Traulsen09} (see Supplementary Material, Section~\ref{SI_Det}). Then, several lineages of mutants can coexist. If $T/2\gg 1/(N\mu_1)$, it is almost certain that some R mutants exist in the population upon the first addition of antimicrobial, which entails $t^\mathrm{a}_\mathrm{R}=T/2$. The horizontal purple solid line plotted at large $T$ in Fig.~\ref{varyT_and_varyN}A, and the horizontal solid lines at large $N$ in Fig.~\ref{varyT_and_varyN}B, both correspond to this deterministic prediction. In the Supplementary Material, Section~\ref{SI_Det}, we study the deterministic limit of our stochastic model, and demonstrate that it matches the results obtained in Fig.~\ref{varyT_and_varyN}A for $N=10^5$ and $N=10^6$ over the whole range of $T$ (see Fig.~\ref{det_vs_stoch}).

The comparison to the spontaneous fitness valley crossing process (Fig.~\ref{varyT_and_varyN}B, black curve and Supplementary Material, Section~\ref{SI_Valley}) demonstrates that periodic alternations of absence and presence of antimicrobial can dramatically accelerate resistance evolution compared to continuous absence of antimicrobial. Recall that within our model, sensitive microorganisms cannot divide with antimicrobial, so resistance cannot evolve at all in continuous presence of antimicrobial. Another possible comparison would be to a continuous presence of a low dose of antimicrobial (below the MIC), but this goes beyond our binary model of antimicrobial action (see Supplementary Information, Section~\ref{SI_Robust} for a discussion of the domain of validity of this model). Alternations are really essential: R mutants appear without antimicrobial, and each addition of antimicrobial rescues the existing R lineages that would be destined to extinction without antimicrobial.

\subsection*{Asymmetric alternations}

We now turn to the more general case of asymmetric alternations of phases of absence and presence of antimicrobial, with respective durations $T_1$ and $T_2$, and $T=T_1+T_2$ (see Fig.~\ref{Fig1}C). 

The average time $t_\mathrm{R}^\mathrm{a}$ when R mutants first exist in the presence of antimicrobial, and start growing, can be obtained by a straightforward generalization of the symmetric alternation case Eq.~\ref{tR}. What matters is how the duration $T_1$ of the phase without antimicrobial, where S individuals can divide and mutate, compares to the average time $\tau_\mathrm{R}^\mathrm{d}$ an R lineage would drift before extinction without antimicrobial. If $T_1\ll\tau_\mathrm{R}^\mathrm{d}$, the first R mutant takes an average time $T/(N\mu_1T_1)$ to appear, and is likely to be rescued by the next addition of antimicrobial. If $T_1\gg\tau_\mathrm{R}^\mathrm{d}$, the fraction of time during which R mutants are present in the phases without antimicrobial is $N\mu_1\tau_\mathrm{R}^\mathrm{d}$, and antimicrobial is added every $T$, so $t_\mathrm{R}^\mathrm{a}=T/(N \mu_1\tau_\mathrm{R}^\mathrm{d})$. Hence, we obtain
\begin{equation}
t_\mathrm{R}^\mathrm{a}=\frac{T}{N \mu_1\,\mathrm{min}(\tau_\mathrm{R}^\mathrm{d},T_1)} \,.
\label{tR_asymm}
\end{equation}

Once the R mutants have taken over the population, the appearance and fixation of C mutants is not affected by the alternations. Hence, Eq.~\ref{TotTime} holds for asymmetric alternations, with $t_\mathrm{R}^\mathrm{a}$ given by Eq.~\ref{tR_asymm}. In the rare mutation regime, if $\mu_1\ll\mu_2$, then $t^\mathrm{f}_\mathrm{C}$ will be dominated by $t^\mathrm{a}_\mathrm{R}$, and if $\mu_1\approx\mu_2$, then $t^\mathrm{f}_\mathrm{C}$ will be dominated by $t^\mathrm{a}_\mathrm{R}$ if $T>\min\left(\tau_\mathrm{R}^\mathrm{d},\,\, T_1\right)/p_\mathrm{RC}$, where $p_\mathrm{RC}$ is the fixation probability of a single C mutant in a population of R individuals.

\begin{figure}[h!]
\centering
  \includegraphics[width=1.0\linewidth]{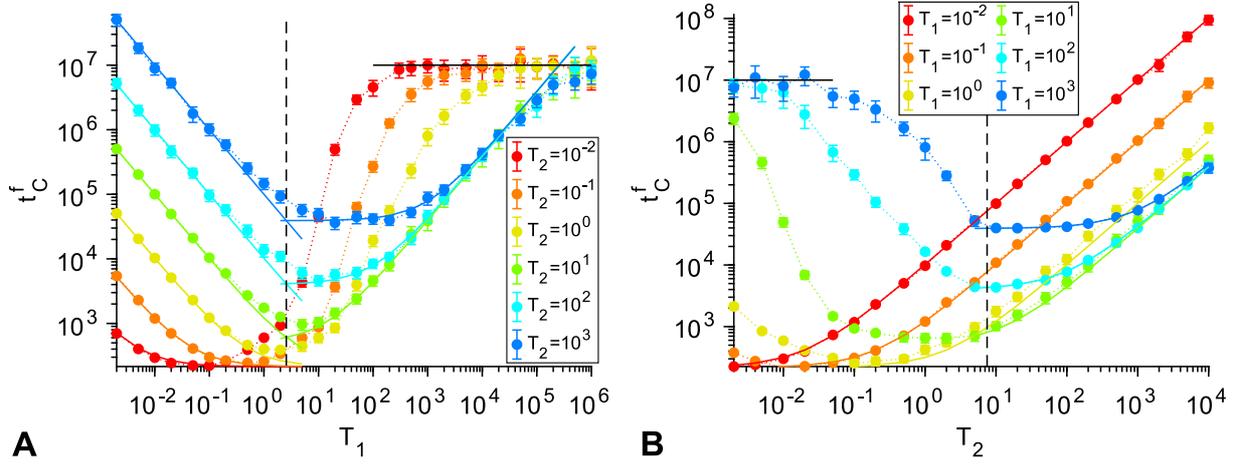}
  \quad
\caption{\textbf{Asymmetric alternations.} Fixation time $t_\mathrm{C}^\mathrm{f}$ of C individuals in a population subjected to asymmetric alternations of absence and presence of antimicrobial (respective durations: $T_1$ and $T_2$). Data points correspond to the average of simulation results (over 10 to $10^3$ replicates), and error bars (sometimes smaller than markers) represent $95\%$ confidence intervals. In both panels, solid lines correspond to our analytical predictions in each regime. In particular, black lines are analytical predictions for fitness valley crossing times in the absence of alternations (see Supplementary Material, Section~\ref{SI_Valley}). Parameter values: $\mu_1=10^{-5}$, $\mu_2=10^{-3}$, $\delta=0.1$ and $N=10^3$. (A) $t_\mathrm{C}^\mathrm{f}$ as function of $T_1$ for different $T_2$. Dashed line: $T_1=\tau_\mathrm{R}^\mathrm{d}$. (B) $t_\mathrm{C}^\mathrm{f}$ as function of $T_2$ for different $T_1$. Dashed line: $T_2=\tau_\mathrm{R}^\mathrm{f}$.}
\label{varyT1_and_varyT2}
\end{figure}

Fig.~\ref{varyT1_and_varyT2}A shows simulation results for $t^\mathrm{f}_\mathrm{C}$ as a function of the duration $T_1$ of the phases without antimicrobial, for different values of the duration $T_2$ of the phases with antimicrobial. As predicted above, we observe a transition at $T_1=\tau_\mathrm{R}^\mathrm{d}$, and different behaviors depending whether $T_2\ll\tau_\mathrm{R}^\mathrm{f}$ or $T_2\gg\tau_\mathrm{R}^\mathrm{f}$. Our analytical predictions from Eqs.~\ref{TotTime} and~\ref{tR_asymm} are plotted in Fig.~\ref{varyT1_and_varyT2}A in the various regimes (solid lines), and are in excellent agreement with the simulation data. The plateau of $t^\mathrm{f}_\mathrm{C}$ at large $T_1$ corresponds to spontaneous valley crossing, and the analytical prediction (see Supplementary Material, Section~\ref{SI_Valley}) is plotted in black in Fig.~\ref{varyT1_and_varyT2}A. 

For $T_2\gg\tau^\mathrm{f}_\mathrm{R}$, Fig. \ref{varyT1_and_varyT2}A shows that $t^\mathrm{f}_\mathrm{C}$ features a striking minimum, which gets higher but wider for longer $T_2$. This can be fully understood from our analytical predictions. Indeed, when $T_1$ is varied starting from small values at fixed $T_2\gg\tau^\mathrm{f}_\mathrm{R}$, different regimes can be distinguished:
\begin{itemize}
\item When $T_1 \ll \tau_\mathrm{R}^\mathrm{d}\left(\lesssim \tau_\mathrm{R}^\mathrm{f}\ll T_2\right)$, Eq.~\ref{tR_asymm} yields $t^\mathrm{a}_\mathrm{R}= T/(N\mu_1T_1)\approx T_2/(N\mu_1T_1)\propto 1/T_1$. 
\item When $\tau_\mathrm{R}^\mathrm{d}\ll T_1 \ll  T_2$, Eq.~\ref{tR_asymm} gives $t^\mathrm{a}_\mathrm{R}=T/(N\mu_1\tau^\mathrm{d}_\mathrm{R})\approx T_2/(N\mu_1\tau^\mathrm{d}_\mathrm{R})$, which is independent from $T_1$.
\item As $T_1$ reaches and exceeds $T_2$, the law $t^\mathrm{a}_\mathrm{R}=T/(N\mu_1\tau^\mathrm{d}_\mathrm{R})$ still holds. It yields $t^\mathrm{a}_\mathrm{R}\approx T_1/(N\mu_1\tau^\mathrm{d}_\mathrm{R})\propto T_1$ when $\tau_\mathrm{R}^\mathrm{d}\ll T_2 \ll  T_1$. 
\end{itemize}
Hence, the minimum of $t^\mathrm{a}_\mathrm{R}$ is $T_2/(N\mu_1\tau^\mathrm{d}_\mathrm{R})\propto T_2$ and is attained for $\tau_\mathrm{R}^\mathrm{d}\ll T_1 \ll  T_2$: it gets higher but wider for larger $T_2$. 

In the opposite regime where $T_2 \ll \tau_\mathrm{R}^\mathrm{d}\lesssim\tau_\mathrm{R}^\mathrm{f}$, Fig.~\ref{varyT1_and_varyT2}A shows that $t^\mathrm{f}_\mathrm{C}$ also features a minimum as a function of $T_1$:
\begin{itemize}
\item When $T_1 \ll T_2\ll \tau_\mathrm{R}^\mathrm{d}$, Eq.~\ref{tR_asymm} yields $t^\mathrm{a}_\mathrm{R}= T/(N\mu_1T_1)\approx T_2/(N\mu_1T_1)\propto 1/T_1$. 
\item When $T_2 \ll T_1\ll \tau_\mathrm{R}^\mathrm{d}$, the same law gives $t^\mathrm{a}_\mathrm{R}= T/(N\mu_1T_1)\approx 1/(N\mu_1)$, which is independent from $T_1$. 
\item When $T_2\ll\tau_\mathrm{R}^\mathrm{d}\ll T_1$, R lineages eventually tend to go extinct, even once they have started growing thanks to an addition of antimicrobial (see Supplementary Material, Section~\ref{SI_Asym} and Fig.~\ref{Fluct_and_ext}B). Then, alternations do not accelerate resistance evolution, and spontaneous valley crossing dominates (black horizontal line in Fig~\ref{varyT1_and_varyT2}A).
\end{itemize}
Hence, the minimum of $t^\mathrm{a}_\mathrm{R}$ is $1/(N\mu_1)$ and is attained for $T_2 \ll T_1\ll \tau_\mathrm{R}^\mathrm{d}$: then, the first R mutant that appears is likely to be rescued by the next addition of antimicrobial, thus driving the complete evolution of resistance in the population. 
For $T_2 \leq T_1\ll \tau_\mathrm{R}^\mathrm{d}$, $t^\mathrm{a}_\mathrm{R}$ is between once and twice this minimum value.

A similar analysis can be conducted if $T_2$ is varied at fixed $T_1$ (Fig.~\ref{varyT1_and_varyT2}B); it is presented in the Supplementary Material, Section~\ref{SI_Asym}. In a nutshell, for asymmetric alternations, a striking minimum for the time of full evolution of resistance by a population occurs when both phases have durations of the same order. Interestingly, the minimum generally occurs when the phases of antimicrobial presence are shorter than those of absence, i.e. $T_2\leq T_1$ (except if $T_2\gg \tau_\mathrm{R}^\mathrm{d}$).

In addition to this minimum, Fig.~\ref{varyT1_and_varyT2} also shows a regime of parameters, when $T_1\ll T_2$ and $T_1\ll\tau_\mathrm{R}^\mathrm{d}$, where the evolution of resistance actually takes longer than fitness valley crossing in the absence of antimicrobial (black lines in Fig.~\ref{varyT1_and_varyT2}). Comparing the timescales involved (see Supplementary Material, Section~\ref{SI_Valley}) shows that in this regime, if $T_2\gg T_1\delta/\mu_2$, the alternation-driven process is faster than the valley-crossing process in the presence of alternations, and thus dominates, but it is slower than the valley-crossing process in the absence of antimicrobial. Hence, in this case, the drug actually slows down the evolution of resistance. Qualitatively, this is because the antimicrobial prevents mutants from arising when it is present.

\section*{Discussion}

\subsection*{Main conclusions}
Because of the generic initial fitness cost of resistance mutations, alternations of phases of absence and presence of antimicrobial induce a dramatic time variability of the adaptive landscape associated to resistance evolution, which alternates back and forth from a fitness valley to an ascending landscape. Using a general and minimal theoretical model which retains the key biological ingredients,  we have shed light on the quantitative implications of these time-varying patterns of selection on the time it takes for resistance to fully evolve \textit{de novo} in a homogeneous microbial population of fixed size. Combining analytical approaches and simulations, we showed that resistance evolution can be driven by periodic alternations of phases of absence and presence of an antimicrobial that stops growth. Indeed, the addition of antimicrobial is able to rescue resistant lineages that were destined to go extinct without antimicrobial.

\begin{figure}[h!]
\centering
\includegraphics[width=1.0\linewidth]{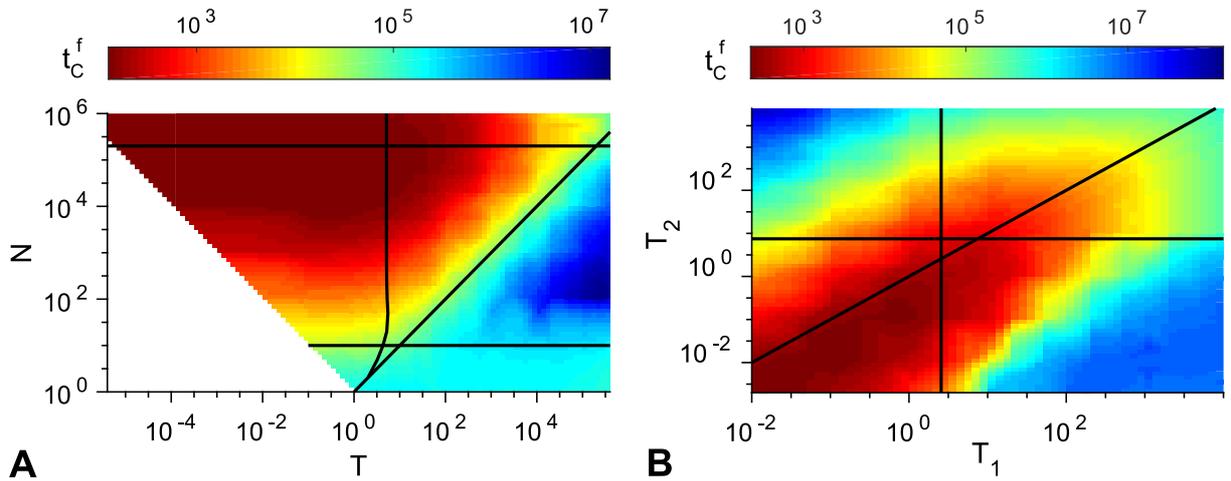}
\quad
\caption{\textbf{Heatmaps.} Fixation time $t_\mathrm{C}^\mathrm{f}$ of C individuals in a population of size $N$ subjected to periodic alternations of absence and presence of antimicrobial. Simulation data plotted in Figs.~\ref{varyT_and_varyN}A and~\ref{varyT1_and_varyT2}A are linearly interpolated. Parameter values: $\mu_1=10^{-5}$, $\mu_2=10^{-3}$, $\delta=0.1$. (A) Symmetric alternations: $t_\mathrm{C}^\mathrm{f}$ as function of the period $T$ and the population size $N$. Top horizontal line: deterministic regime limit $N=1/\mu_1$. Bottom horizontal line: neutral regime limit $N=1/\delta$. Quasi-vertical curve: $T=2\tau_\mathrm{R}^\mathrm{d}$. Diagonal line: $T=N$. Note that no data is shown for $T/2<1/N$ because of the discreteness of our model, which can only deal with timescales larger or equal to the duration of one Moran step, i.e. $1/N$ generation. (B) Asymmetric alternations: $t_\mathrm{C}^\mathrm{f}$ as function of the durations $T_1$ and $T_2$ of the phases of absence and presence of antimicrobial. Vertical line: $T_1=\tau_\mathrm{R}^\mathrm{d}$. Horizontal line: $T_2=\tau_\mathrm{R}^\mathrm{f}$. Diagonal line: $T_1=T_2$. Here $N=10^3$, so the first resistant mutant appears after an average time $T/(N\mu_1 T_1)=10^2\,T/T_1$.  }
  \label{heat}
\end{figure}

We found that fast alternations strongly accelerate the evolution of resistance. In the limit of short alternation periods, the very first resistant mutant that appears is likely to ultimately lead to full resistance of the population, as it will generally be rescued by the next addition of antimicrobial before going extinct, which would be its most likely fate without antimicrobial. For larger periods $T$, the time needed for resistance to evolve increases linearly with $T$, until it reaches the spontaneous valley-crossing time with alternations, which constitutes an upper bound. Our complete stochastic model allowed us to investigate the impact of population size $N$, beyond the limit $N \gg 1/\mu_1$ addressed by deterministic models. We showed that the acceleration of resistance evolution is stronger for larger populations, eventually reaching a plateau in the deterministic limit. Over a large range of intermediate parameters, the time needed for the population to fully evolve resistance scales as $T/N$. These results are summed up in Fig.~\ref{heat}A. 

For asymmetric alternations, featuring different durations $T_1$ and $T_2$ of the phases of absence and presence of antimicrobial, we have shed light on the  existence of a minimum for the time  taken by the population to fully evolve resistance. This striking minimum occurs when both phases have durations of the same order, generally with $T_1\leq T_2$. Moreover, the minimum value reached for the time of resistance evolution decreases for shorter alternation periods. These results are summed up in Fig.~\ref{heat}B.  

\subsection*{Context and perspectives}

Our approach is complementary to previous studies providing a detailed modeling of specific treatments~\cite{Lipsitch97, Wahl00,Regoes04,Wu14,Meredith15,AbelzurWiesch14,Ke15}. Indeed, the majority of them~\cite{Lipsitch97,SchulzzurWiesch10, Wahl00,Regoes04,Wu14,Meredith15,Bauer17} neglect stochastic effects, while they can have a crucial evolutionary impact~\cite{Fisher07,Ewens79}. The deterministic approach is appropriate if the number $N$ of competing microbes satisfies $N\mu_1\gg 1$, where $\mu_1$ is the mutation rate~\cite{Rouzine01,Fisher07}. Such large sizes can be reached in some established infections~\cite{Hughes15}, but microbial populations go through very small bottleneck sizes (sometimes $N\sim 1 - 10$~\cite{Gutierrez12}) when an infection is transmitted. Moreover, established microbial populations are structured, even within a single patient~\cite{VanMarle07}, and competition is local, which decreases the effective value of $N$. Some previous studies did take stochasticity into account, but several did not include compensation of the cost of resistance~\cite{NissenMeyer66,Hansen17}, while others made specific epidemiological assumptions~\cite{AbelzurWiesch14}. 

Given the usual steepness  of pharmacodynamic curves~\cite{Regoes04}, we have modeled the action of a biostatic antimicrobial in a binary way, with no growth inhibition under the MIC and full growth inhibition of S microorganisms above it (see Model). An analysis of the robustness of this approximation is presented in the Supplementary Material, Section~\ref{SI_Robust}, showing that it is appropriate if the rise time, i.e. the time needed for the fitness of sensitive microorganisms to switch from a low value to a high value and vice-versa when antimicrobial is removed or added, is short enough (see Fig.~\ref{robustness}). Qualitatively, if this rise time is shorter than the other environmental and evolutionary timescales at play, then the fitness versus time function is effectively binary.

Our model assumes that the size of the microbial population remains constant. While this is realistic in some controlled experimental setups, e.g. turbidostats~\cite{Myers44}, microbial populations involved in infections tend to grow, starting from a small transmission bottleneck, and the aim of the antimicrobial treatment is to make them decrease in size and eventually go extinct. In the case of biostatic antimicrobials, which prevent bacteria from growing, populations can go extinct due to spontaneous and immune system-induced death. Our model with constant population size should however be qualitatively relevant at the beginning and middle stages of a treatment (i.e. sufficiently after transmission and before extinction). Constant population sizes facilitate analytical calculations, and allowed us to fully quantify the impact of a periodic presence of antimicrobial on resistance evolution, but it will be very interesting to extend our work to variable population sizes~\cite{Melbinger10,Melbinger15,Huang15}. This would allow us to model biocidal antimicrobials, and to include effects such as antibiotic tolerance, which tend to precede resistance under intermittent antibiotic exposure~\cite{Levin-Reisman17}. Another exciting extension would be to incorporate spatial structure~\cite{Bitbol14,Nahum15,Cooper15} and environment heterogeneity, in particular drug concentration gradients. Indeed, static gradients can strongly accelerate resistance evolution~\cite{Zhang11,Greulich12,Hermsen12,Baym16}, and one may ask how this effect combines with the temporal alternation-driven one investigated here. Besides, it would be interesting to compare the impact of periodic alternations to that of random switches of the environment~\cite{Mustonen08,Rivoire11,Melbinger15,Desponds16,Wienand17}.

\subsection*{Implications for clinical and experimental situations}

The situation where the phases of absence and presence of antimicrobial have similar durations ($T_1\approx T_2$) yields a dramatic acceleration of resistance evolution, and is unfortunately clinically realistic. Indeed, a goal in treatment design is that the serum concentration of antimicrobial exceeds the MIC for at least 40 to 50\% of the time~\cite{Jacobs01}, which implies that actual treatments may involve the alternations that most favor resistance evolution according to our results~\cite{Jacobs01,Regoes04}. Besides, bacteria divide on a timescale of about an hour (yielding a $\tau_\mathrm{R}^\mathrm{d}$ of order of a few hours), and antimicrobial is often taken every 8 to 12 hours in treatments by the oral route, so the alternation period does not last for many generations: this is close to our worst-case scenario of short symmetric periods.

In this worst case scenario, full \textit{de novo} resistance evolution can result from the appearance of the very first R mutant, which takes $T/(N\mu_1 T_1)$. Indeed, its lineage is likely to be rescued by the next addition of antimicrobial. Under the conservative assumption that only one resistance mutation is accessible, taking $\mu_1\sim 10^{-10}$, which is the typical mutation probability per nucleotide and per generation in \textit{Escherichia coli} bacteria~\cite{Wielgoss11}, and taking $\delta\sim 0.1$~\cite{Schrag97}, we find that this duration is less than a day ($\sim 10-20$ generations) for $N\sim 10^9$, and a few days for $N\sim 10^8$, numbers that can be reached in infections~\cite{Hughes15,zurWiesch11}. For such large populations, the fixation of the C (compensated) mutant will take more time, but once R is fixed (which takes $\sim\!1$ day after the appearance of the first R mutant), C is very likely to fix even if the treatment is stopped. This is due to the large number of compensatory mutations, which yields a much higher effective mutation rate toward compensation than toward reversion to sensitivity~\cite{Levin00,Paulander07,Hughes15}. In addition, many mutations to resistance are often accessible, yielding higher effective $\mu_1$, e.g. $\mu_1\sim10^{-8}$ for rifampicin resistance in some wild isolates of \textit{E. coli}~\cite{deSousa15}, meaning that smaller populations can also quickly become resistant in the presence of alternations. Recall that we are only considering \textit{de novo} resistance evolution, without pre-existent resistant mutants, or other possible sources of resistance, such as horizontal gene transfer, which would further accelerate resistance acquisition. 

In summary, an antimicrobial concentration that drops below the MIC between each intake can dramatically favor \textit{de novo} resistance evolution. More specifically, we showed that the worst case occurs when $T_1\leq T_2$, which would be the case if the antimicrobial concentration drops below the MIC relatively briefly before each new intake. Our results thus emphasize how important it is to control for such apparently innocuous cases, and constitute a striking argument in favor of the development of extended-release antimicrobial formulations~\cite{Gao11}. 

While the parameter range that strongly accelerates resistance evolution should preferably be avoided in clinical situations, it could be tested and harnessed in evolution experiments. Again, these parameters are experimentally accessible. Controlled variations of antimicrobial concentration are already used experimentally, in particular in morbidostat experiments~\cite{Toprak11}, where the population size is kept almost constant, which matches our model. In Ref.~\cite{Toprak11}, a dramatic and reproducible evolution of resistance was observed in $\sim 20$ days when periodically adjusting the drug concentration to constantly challenge \textit{E. coli} bacteria. Given our results, it would be interesting to test whether resistance evolution could be made even faster by adding drug in a turbidostat with a fixed periodicity satisfying $T_1\leq T_2\ll \tau_\mathrm{R}^\mathrm{d}$.

\section*{Acknowledgments}
We thank Claude Loverdo, David J. Schwab and Raphaël Voituriez for stimulating discussions. AFB also acknowledges the KITP Program on Evolution of Drug Resistance (KITP, Santa Barbara, CA, 2014), which was supported in part by the National Science Foundation under Grant NSF PHY 17-48958. LM acknowledges funding by a graduate fellowship from EDPIF.

\newpage

\beginsupplement

%%%%

\begin{center}
\LARGE{\textbf{Supplementary Material}}
\end{center}

\vspace{0.1cm}

\section{Table of notations}

\begin{table}[h!]
\centering
\begin{tabular}{|p{0.1\linewidth}|p{0.85\linewidth}|}
  \hline
  Notation & Definition \\
\hline\hline
  S & Sensitive microorganisms \\
    \hline
  R & Resistant microorganisms \\
    \hline
  C & Resistant-compensated microorganisms \\
      \hline
$T$ & Period of the alternations of absence and presence of antimicrobial \\
    \hline
     $T_1$ & Duration of the phase without antimicrobial (for asymmetric alternations)\\
      \hline
    $T_2$ & Duration of the phase with antimicrobial (for asymmetric alternations) \\
  \hline
  $N$ & Population size \\
    \hline
    $\delta$ & Fitness cost of antimicrobial resistance\\
  \hline
  $\mu_1$ & Mutation rate from S to R \\
    \hline
  $\mu_2$ & Mutation rate from R to C \\
    \hline
  $t_{\mathrm{C}}^{\mathrm{f}}$ & Total time of full resistance evolution (time until the C type fixes, starting from a population of S individuals)\\
    \hline
  $t_{\mathrm{R}}^{\mathrm{a}}$ & Average time when R individuals first exist in the presence of antimicrobial, starting from a population of S individuals\\
    \hline
  $t_{\mathrm{C}}^{\mathrm{a}}$ & Average time when the first C mutant whose lineage will fix appears, starting from a population of R individuals\\
         \hline
  $\tau_{\mathrm{R}}^{\mathrm{d}}$ & Average lifetime of the lineage of a single R mutant, until it disappears, in a population of S individuals, in the absence of antimicrobial\\
     \hline
  $\tau_{\mathrm{R}}^{\mathrm{f}}$ & Average fixation time of the lineage of a single R mutant in a population of S individuals, in the presence of antimicrobial \\
    \hline
      $\tau_{\mathrm{C}}^{\mathrm{f}}$ & Average fixation time of the lineage of a single C mutant in a population of R individuals\\
\hline
  $p_{\mathrm{SR}}$ & Fixation probability of a single R mutant in a population of S individuals in the absence of antimicrobial\\
                    \hline
         $p_{\mathrm{RC}}$ & Fixation probability of a single C mutant in a population of R individuals \\
      \hline
\end{tabular}
\vspace{0.5cm}
\caption{\textbf{Notations.} This table lists the different notations introduced in the main text and their meaning.}
\label{Not_table}
\end{table}

\section{Fixation probabilities and fixation times in the Moran process}
\label{SI_Moran}

Here, we discuss in detail the fixation probabilities and mean fixation times in the Moran process, which are used throughout the main text. These quantities are already known~\cite{Ewens79,Traulsen09}, but we present a derivation for the sake of pedagogy and completeness. Our derivation is based on the general formalism of first passage times, and gives the same results as those obtained in the literature, often using other methods~\cite{Ewens79,Traulsen09}. Next, we use the general expressions obtained to express the various fixation probabilities and fixation times used in the main text.

\subsection{The Moran process}

The Moran model \cite{Moran58,Ewens79} is a simple stochastic process used to describe the evolution of the composition of asexual populations of finite and constant size.  It allows one to incorporate  variety-increasing processes such as mutation and variety-reducing processes such as natural selection. 

\begin{figure}[h!]
\begin{center}
   \includegraphics[width=0.95\textwidth]{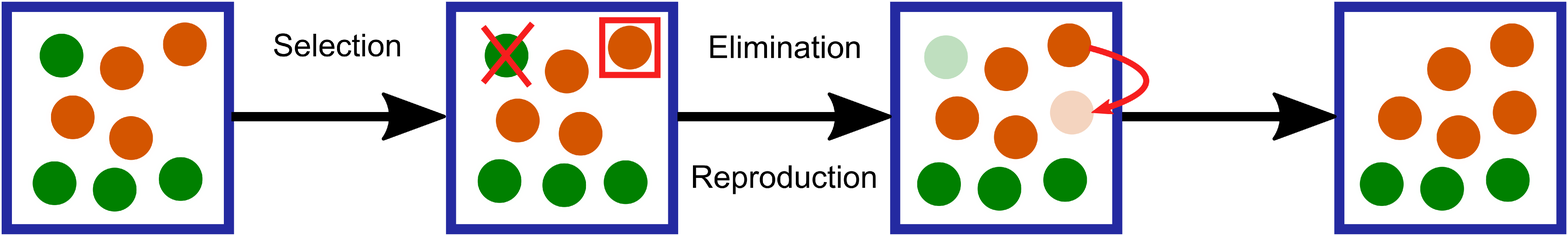}
   \quad
   \caption{\textbf{Sketch of the Moran process.} One step of the Moran process is represented in a population with 8 individuals of 2 different types (different colors). }
   \label{MP}
\end{center}
\end{figure}

In the Moran model, at each time step, an individual is chosen at random to reproduce and another one is chosen to die (see Figure \ref{MP}). Hence, the total number of individuals in the population stays constant. Note that we will consider that the same individual can be selected to reproduce and die at the same step. Natural selection can be introduced by choosing the individual that reproduces with a probability proportional to its fitness. To implement mutations upon division, one can allow the offspring to switch type with a certain probability at each step. 
When a mutant arises within the Moran model at constant fitness, its lineage can either disappear or fix in the population, i.e. take over the whole population. The outcome is not fully determined by fitness differences as in a deterministic case, but also by stochastic fluctuations, also known as genetic drift. Here, we focus on the evolution of population composition under genetic drift and selection alone. In the rare mutation regime, these processes are much faster than the time between the occurrence of two mutations, so mutation can be neglected during the process of fixation of one type.  The Moran model allows us to compute  explicit expressions for quantities such as fixation probabilities and fixation times \cite{Ewens79,Taylor06} (see below).

Let us consider a population of $N$ individuals of two types A and B, which have fitnesses $f_A$ and $f_B$, respectively. We denote the number of A individuals by $j$. Thus $N-j$ represents the number of B individuals. Let us study the evolution of $j$ at one step of the Moran process (for an example, see Figure~\ref{MP}). The transition probabilities associated to the Moran process read~\cite{Ewens79}:

\begin{equation}
\begin{cases}
\displaystyle \Pi_{j \rightarrow j+1}=\frac{N-j}{N}\frac{f_A j}{f_A j +f_B (N-j)} \\
\displaystyle \Pi_{j \rightarrow j-1}=\frac{j}{N}\frac{f_B(N-j)}{f_A j + f_B (N-j)} \\
\displaystyle \Pi_{j \rightarrow j}=1-\Pi_{j \rightarrow j+1}-\Pi_{j \rightarrow j-1}\mbox{ }.
\end{cases}
\label{trans_rates}
\end{equation}

The Moran process is a discrete-time Markov process, since the probabilities of states $j$ after one step only depend upon the present value of $j$. Let us take the limit of continuous time and write the master equation $\dot{{\bf P}}={\bf R_A}{\bf P}$ giving the probability of being at state $j$ at time $t$:

\begin{equation}
\frac{d}{dt}\begin{pmatrix}
P_0 \\
P_1 \\
P_2 \\
\vdots \\
P_N
\end{pmatrix}=
\begin{pmatrix}
-\Pi_{0 \rightarrow 1} & \Pi_{1 \rightarrow 0} & 0 & \cdots & 0\\
\Pi_{0 \rightarrow 1} & -(\Pi_{1 \rightarrow 0}+\Pi_{1 \rightarrow 2}) & \Pi_{2 \rightarrow 1} & (0) & \vdots\\
0 & \Pi_{1 \rightarrow 2} & -(\Pi_{2 \rightarrow 1}+\Pi_{2 \rightarrow 3}) & \ddots & 0\\
\vdots & (0) & \ddots & \ddots & \Pi_{N \rightarrow N-1} \\
0 & \cdots & 0 & \Pi_{N-1 \rightarrow N} & -\Pi_{N \rightarrow N-1}
\end{pmatrix}
\begin{pmatrix}
P_0 \\
P_1 \\
P_2 \\
\vdots \\
P_N
\end{pmatrix}\mbox{ }.
\label{Master}
\end{equation}
This Markov chain has two absorbing states, namely $j=0$ and $j=N$, which correspond to the fixation of B and A individuals, respectively. Once these states are reached, no more changes can occur, in the absence of mutation. It follows that all the components of the first and the last columns of ${\bf R_A}$ equal to 0 (see Eq.~\ref{Master}), so ${\bf R_A}$ is not invertible. In the following, we will denote by ${\bf \widetilde{R}_A}$ the reduced transition rate matrix in which the rows and the columns corresponding to the absorbing states ($j=0$, $j=N$) are removed, and by ${\bf \widetilde{R}_A}^{-1}$ its inverse. Let us note that ${\bf R_A}$ is a tridiagonal matrix, which allows for major simplifications of analytical calculations \cite{Ewens79}. Note that in order to obtain the transition rate matrix associated to B individuals, one just needs to apply the reversal $j \leftrightarrow N-j$. This corresponds to using the matrix ${\bf R_B}={\bf J}{\bf R_A}{\bf J}$ where ${\bf J}$ is the anti-identity matrix. For instance, in $2$ dimensions, ${\bf J}=\begin{pmatrix}
0 & 1 \\
1 & 0 
\end{pmatrix}$.

\subsection{General fixation probabilities and fixation times}

\paragraph*{Definitions.}

The fixation probability $\phi^A_{j_0}$ represents the probability that A individuals finally succeed and take over the population, starting from $j=j_0$ individuals of type A. In particular, $\phi_0^A=0$ and $\phi_N^A=1$. Similarly, $\phi^B_{j_0}$ is the fixation probability of the B individuals, still starting from $j=j_0$ individuals of type A.

Mean fixation times are the mean times to reach one of the absorbing states. The unconditional fixation time $t_{j_0}$ is the average time until fixation in either $j=0$ or $j=N$, when starting from a number $j=j_0$ of A individuals. The conditional fixation time $t_{j_0}^A$ corresponds to the average time until fixation in $j=N$, when starting from $j_0$, provided that type A fixes. Note that in what follows, we will express the fixation times in numbers of steps of the Moran process. Conversion to generations can then be performed by dividing the number of Moran steps by $N$.

In the following, we present a derivation of the fixation probabilities and of the fixation times in the Moran process~\cite{Ewens79,Traulsen09} that uses the general formalism of mean first passage times \cite{Sekimoto10}.  

\paragraph*{Fixation probabilities.}
Assuming that at $t=0$, the system is at state $j=j_0$, let us focus on the fixation probability $\phi_{j_0}^{A}$ of the A type in the population. The stochastic process stops at the time $\widehat{\tau}_{FP}$ when $j$ fixes, i.e. first reaches one of the absorbing states $\{j=0,\,j=N\}$. Hence, integrating over all values of $\widehat{\tau}_{FP}$, under the condition that fixation finally occurs in $j=N$, yields
\begin{equation}
\phi_{j_0}^A=\int_{0}^{\infty}p(\widehat{\tau}_{FP} \in [t,t+dt] \: | \: j_0 \, ,j_{\infty}=N)=\Pi_{N-1 \rightarrow N} \int_{0}^{\infty}P_{N-1}(t)dt\,.
\end{equation}
In the last expression, we have taken advantage of the fact that the only way to fix in $j=N$ between $t$ and $t+dt$ is to be in state $j=N-1$ at time $t$ and then to transition from $N-1$ to $N$ (see Eq.~\ref{Master}). We have thus introduced the probability $P_{N-1}(t)$ of being in state $j=N-1$ at time $t$, starting in state $j=j_0$ at time 0. More generally, the probability $P_i(t)$ can be considered.

Integrating the Master equation Eq.~\ref{Master} to determine $P_i(t)$, with the initial condition $P_i(0)=\delta_{i\,j_0}$, where $\delta_{i \, j_0}$ denotes the Kronecker delta, which is equal to 1 if $i=j_0$ and 0 otherwise, yields
\begin{equation}
\phi_{j_0}^A=-\Pi_{N-1 \rightarrow N}({\bf \widetilde{R}_A}^{-1})_{N-1 \, j_0}\,.
\label{proba_A}
\end{equation}

A similar reasoning gives the fixation probability $\phi_{j_0}^B$ of the B type, still starting from $j_0$ individuals of type A and $N-j_0$ individuals of type B:
\begin{equation}
\phi_{j_0}^B=-\Pi_{1 \rightarrow 0}({\bf \widetilde{R}_A}^{-1})_{1 \, j_0}\,.
\label{proba_B}
\end{equation}
These two probabilities satisfy $\phi_{j_0}^A+\phi_{j_0}^B=1$ since there are 2 absorbing states in the process. 

\paragraph*{Mean fixation times.} 
Let us now focus on the mean fixation times, still assuming that at $t=0$, the system is at state $j=j_0$. The probability that fixation in one of the absorbing states $\{j=0,j=N\}$ occurs between $t$ and $t+dt$ reads:
\begin{equation}
p(\widehat{\tau}_{FP} \in [t,t+dt] \: | \: j_0)=\sum_{i=1}^{N-1}P_i(t)-\sum_{i=1}^{N-1}P_i(t+dt)= -\sum_{i=1}^{N-1}\frac{dP_i}{dt}dt\,,
\label{proba_1st}
\end{equation}
where, as above, $P_i(t)$ represents the probability of being in state $i$ at time $t$ starting in $j_0$ at time 0 (note that the initial condition $j_0$ is omitted for brevity). Thus, the unconditional fixation time can be expressed as:
\begin{align}
t_{j_0}&=\mathbb{E}[\widehat{\tau}_{FP}\: |\: j_0] =\int_{0}^{\infty}t\, p(\widehat{\tau}_{FP} \in [t,t+dt] \: | \: j_0)\\
&=-\sum_{i=1}^{N-1} \int_{0}^{\infty}t\, \frac{dP_i}{dt}\,dt = \sum_{i=1}^{N-1} \int_{0}^{\infty}P_i(t)\,dt \mbox{ }.
\end{align}
Here, we used Eq.~\ref{proba_1st}, where the sums run over all the states that are not absorbing ($1\leq i\leq N-1$). We also performed an integration by parts, and used $[t \,P_i(t)]_0^\infty=0$ for $1\leq i\leq N-1$, which holds because the probability of reaching an absorbing state of the Markov chain tends to 1 as $t \rightarrow \infty$. Integrating the Master equation Eq.~\ref{Master} to determine $P_i(t)$, with the initial condition $P_i(0)=\delta_{i\,j_0}$, gives
\begin{equation}
\displaystyle t_{j_0} = -\sum_{i=1}^{N-1} ({\bf \widetilde{R}_A}^{-1})_{i\,j_0} \mbox{ }.
\label{t_unc}
\end{equation}

To express the conditional fixation time $t_{j_0}^A$ of type A, starting from $j_0$ A individuals, we need to take into account the condition that fixation finally occurs in state $j=N$:
\begin{equation}
p(\widehat{\tau}_{FP} \in [t,t+dt] \: | \: j_0, j_\infty=N)=\sum_{i=1}^{N-1}p(i \: | \: j_0,j_\infty=N)(t)- \sum_{i=1}^{N-1}p(i \: | \: j_0,j_\infty=N)(t+dt) \mbox{ }.
\end{equation}
The Bayes relation gives:
\begin{equation}
p(j \: | \: j_0,j_\infty=N)=\frac{\phi_j^A}{\phi_{j_0}^A}P_j \mbox{ }.
\end{equation}
By using the same method as for the unconditional fixation time, one obtains:
\begin{equation}
\displaystyle t_{j_0}^A=-\frac{1}{\phi_{j_0}^A}\sum_{i=1}^{N-1}\phi_i^A ({\bf \widetilde{R}_A}^{-1})_{i \, j_0}\mbox{ }.
\label{t_N-1_A}
\end{equation}
Similarly, the conditional fixation time of the B type, starting from $j_0$ A individuals, reads:
\begin{equation}
\displaystyle t_{j_0}^B=-\frac{1}{\phi_{j_0}^B}\sum_{i=1}^{N-1}\phi_{N-i}^B({\bf \widetilde{R}_B}^{-1})_{i \, N-j_0}\mbox{ }.
\label{t_1_B}
\end{equation}
It is straightforward to verify that Eqs. \ref{t_unc}, \ref{t_N-1_A} and \ref{t_1_B} are linked by the relation:
\begin{equation}
t_{j_0}=\phi_{j_0}^B t_{j_0}^B + \phi_{j_0}^A t_{j_0}^A
\label{all_times} \mbox{ }.
\end{equation}

\paragraph*{Neutral drift.}

Let us first consider the case without selection $f_A=f_B$. In this case, the Moran process can be seen as a non-biased random walk, since individuals of both types are equally likely to be picked for reproduction and death. Fixation eventually happens due to fluctuations. This process, called neutral drift \cite{Ewens79} corresponds to diffusion in physics. The transition rates of the system (\ref{trans_rates}) simplify as follows:
\begin{equation}
\begin{cases}
\displaystyle \Pi_{j \rightarrow j+1}=\Pi_{j \rightarrow j-1}=\frac{j(N-j)}{N^2} \\
\displaystyle \Pi_{j \rightarrow j}=1-2\frac{j(N-j)}{N^2}\mbox{ }.
\end{cases}
\label{trans_rates_neutral}
\end{equation}
Note that here, $j$  can denote the number of A or B individuals indifferently. Indeed, the symmetry $j \leftrightarrow N-j$ entails ${\bf R_A}={\bf R_B}={\bf R}$, and the transition rate matrix is centrosymmetric, i.e. ${\bf R}={\bf J}{\bf R}{\bf J}$. For consistency, we will continue to call $j$ the number of A individuals.

The fixation probability $\phi_{j_0}^A$ can be obtained from  Eq. \ref{proba_A}. It involves elements of the inverse of the transition rate matrix. Solving ${\bf \widetilde{R}}{\bf \widetilde{R}}^{-1}={\bf I}$, where ${\bf I}$ is the identity matrix, gives
\begin{equation}
({\bf \widetilde{R}}^{-1})_{N-1 \, i}=-\frac{i \, N}{N-1}\mbox{ for } 1 \leq i \leq N-1\,.
\label{inv_mat_neutral}
\end{equation}
Hence, 
\begin{equation}
\phi_{j_0}^A=\frac{j_0}{N}\,.
\label{proba_A_neutral}
\end{equation}
Taking advantage of the centrosymmetry of ${\bf{R}}$ (see above), a property which transfers to ${\bf \widetilde{R}}$ and ${\bf \widetilde{R}}^{-1}$, and entails $({\bf \widetilde{R}}^{-1})_{1 \, j_0}=({\bf \widetilde{R}}^{-1})_{N-1 \, N-j_0}$, we can apply Eq.~\ref{proba_B}, yielding
\begin{equation}
\phi_{j_0}^B=\frac{N-j_0}{N}\,.
\label{proba_B_neutral}
\end{equation}
Note that $\phi_{j_0}^A+\phi_{j_0}^B=1$, as expected.

Let us now express the fixation times, focusing on the fate of a single mutant of type B, which corresponds to $j_0=N-1$. To compute the unconditional fixation time $t_{N-1}$, we again need elements of the inverse of the transition rate matrix (see Eq. \ref{t_unc}), which are given by
\begin{equation}
({\bf \widetilde{R}}^{-1})_{i \, N-1}=-\frac{N}{N-i}\mbox{ }.
\label{inv_mat_neutral_2}
\end{equation}
Using Eqs. \ref{t_unc} and \ref{inv_mat_neutral_2}, we obtain:
\begin{equation}
\displaystyle t_{N-1}=N \sum_{i=1}^{N-1}\frac{1}{i}\mbox{ }.
\label{t_N-1_neutral}
\end{equation}
Similarly, using Eqs. \ref{t_N-1_A}, \ref{proba_A_neutral} and~\ref{inv_mat_neutral_2}, we obtain the conditional fixation time of type A:
\begin{equation}
\displaystyle t_{N-1}^A=\frac{N^2}{N-1}\sum_{i=2}^{N}\frac{1}{i}\mbox{ }.
\label{t_N-1_A_neutral}
\end{equation}
Finally, using Eqs. \ref{t_1_B}, \ref{proba_A_neutral} and~\ref{inv_mat_neutral_2}, and making use of the centrosymmetry of ${\bf \widetilde{R}}^{-1}$ (see above), yields the conditional fixation time of type B:
\begin{equation}
t_{N-1}^B=N(N-1)\mbox{ }.
\label{t_N-1_B_neutral}
\end{equation}

\paragraph*{Selection.}

Let us now study the more general case involving selection. For this, let us consider two types A and B having different fitnesses $f_A$ and $f_B$, and let us introduce $\gamma=f_A/f_B$. Note that with selection, the transition rate matrices ${\bf R_A}$ and ${\bf R_B}={\bf J}{\bf R_A}{\bf J}$ are different.
In order to compute the fixation probability $\phi_{j_0}^A$, we need some elements of the inverse of the transition rate matrix ${\bf \widetilde{R}_A}^{-1}$, which are given by:
\begin{equation}
({\bf \widetilde{R}_A}^{-1})_{N-1 \, i}=-\frac{N}{N-1}\frac{1-\gamma^{-i}}{1-\gamma^{-N}}\left(N-1+\gamma^{-1}\right)\mbox{ for }1 \leq i \leq N-1 \,.
\end{equation}
Then, using the previous result and Eq. \ref{proba_A}, one obtains:
\begin{equation}
\phi_{j_0}^A=\frac{1-\gamma^{-j_0}}{1-\gamma^{-N}}\mbox{ },
\label{fix_prob_A}
\end{equation}
and $\phi_{j_0}^A+\phi_{j_0}^B=1$ yields:
\begin{equation}
\phi_{j_0}^B=\frac{1-\gamma^{N-j_0}}{1-\gamma^{N}}\mbox{ }.
\label{fix_prob_B}
\end{equation}

Let us now turn to the fixation times. According to Eq. \ref{t_unc}, we need to compute other elements of the inverse of the transition rate matrix ${\bf \widetilde{R}_A}^{-1}$. Those satisfy:
\begin{equation}
({\bf \widetilde{R}_A}^{-1})_{i \, N-1}=\frac{N}{i(N-i)}\frac{1-\gamma^i}{1-\gamma^N}\left(i-i\gamma-N\right)\textrm{ for $1\leq i\leq N-1$}.
\label{inv_RA}
\end{equation}
Using Eqs. \ref{t_unc} and~\ref{inv_RA}, the unconditional fixation time reads:
\begin{equation}
t_{N-1}=\frac{N}{1-\gamma^N}\sum_{i=1}^{N-1}\frac{(N+i\gamma-i)(1-\gamma^i)}{i(N-i)}\,.
\label{t_n-1_select}
\end{equation}

To compute the conditional fixation time $t_{N-1}^A$, we substitute Eqs. \ref{fix_prob_A} and \ref{inv_RA} in Eq. \ref{t_N-1_A}, obtaining:
\begin{equation}
t_{N-1}^A=\frac{N}{(1-\gamma^N)(1-\gamma^{1-N})}\sum_{i=1}^{N-1}\frac{(N+i\gamma-i)(1-\gamma^i)(1-\gamma^{-i})}{i(N-i)}\,.
\label{t_N-1_A_select}
\end{equation}

A similar reasoning can be used to obtain the conditional fixation time $t_{N-1}^B$ starting from Eq. \ref{t_1_B}. In order to express the required $({\bf \widetilde{R}_B}^{-1})_{j \, 1}$, we combine the relation ${\bf \widetilde{R}_B}={\bf J}{\bf \widetilde{R}_A}{\bf J}$, which implies ${\bf \widetilde{R}_B}^{-1}={\bf J}{\bf \widetilde{R}_A}^{-1}{\bf J}$, together with Eq.~\ref{inv_RA}, and obtain
\begin{equation}
({\bf \widetilde{R}_B}^{-1})_{i \, 1}=\frac{N}{i(N-i)}\frac{1-\gamma^{N-i}}{1-\gamma^N}\left(i\gamma-i-N\gamma\right) \textrm{ for $1\leq i\leq N-1$.}
\label{inv_RB}
\end{equation}
This finally yields
\begin{equation}
t_{N-1}^B=\frac{N}{(1-\gamma^N)(1-\gamma)}\sum_{i=1}^{N-1}\frac{(N+i\gamma-i)(1-\gamma^i)(1-\gamma^{N-i})}{i(N-i)}\,.
\label{t_1_R_select}
\end{equation}

\subsection{Fixation probabilities and fixation times used in the main text}

Let us now make an explicit link between the general expressions obtained above and the fixation probabilities and fixation times used in the main text.

\paragraph*{Fixation probabilities.} First, in the main text, $p_{\mathrm{SR}}$ represents the probability that a single resistant (R) mutant fixes without antimicrobial in a population of size $N$ where all other individuals are of type S. Without antimicrobial, $f_\mathrm{S}=1$ and $f_\mathrm{R}=1-\delta$. Considering S as type A and R as type B, we have $\gamma=f_\mathrm{S}/f_\mathrm{R}=1/(1-\delta)$, and our initial condition is $j_0=N-1$. Hence, Eq. \ref{fix_prob_B} yields
\begin{equation}
p_{\mathrm{SR}}=\phi_{N-1}^\mathrm{R} =\frac{1-\left(1-\delta\right)^{-1}}{1-\left(1-\delta\right)^{-N}}\,.
\label{pSR}
\end{equation}
In particular, in the effectively neutral case where $\delta \ll 1$ and $N\delta \ll 1$, it yields
\begin{equation}
p_{\mathrm{SR}}\approx\frac{-\delta}{1-e^{-N\log(1-\delta)}}\approx\frac{-\delta}{1-e^{N\delta}}\approx\frac{1}{N}\,,
\end{equation}
i.e. we recover the result of the neutral case $\delta=0$ (see Eq. \ref{proba_A_neutral}). Conversely, in the regime where $\delta \ll 1$ and $N\delta \gg 1$, Eq.~\ref{pSR} yields
\begin{align}
p_{\mathrm{SR}}&\approx\frac{-\delta}{1-e^{N\delta}} \approx\delta e^{-N \delta} \mbox{ }.
\end{align}

Second, $p_{\mathrm{RC}}$ denotes the fixation probability of a single C individual in a population of size $N$ where all other individuals are of type R. Independently of antimicrobial presence, $f_\mathrm{R}=1-\delta$ and $f_\mathrm{C}=1$. Considering R as type A and C as type B, we have $\gamma=f_\mathrm{R}/f_\mathrm{C}=1-\delta$, and our initial condition is $j_0=N-1$. Hence, Eq. \ref{fix_prob_B} yields
\begin{equation}
p_{\mathrm{RC}}=\phi_{N-1}^\mathrm{C}=\frac{\delta}{1-\left(1-\delta\right)^N}\,.
\label{pRC}
\end{equation}
In particular, in the effectively neutral case where $\delta \ll 1$ and $N\delta \ll 1$, it yields
\begin{equation}
p_{\mathrm{RC}}=\frac{\delta}{1-e^{N\log (1-\delta)}}\approx \frac{\delta}{1-e^{-N \delta}}\approx\frac{1}{N}\,,
\end{equation}
i.e. we again recover the result of the neutral case $\delta=0$ (see Eq. \ref{proba_A_neutral}). Conversely, in the regime where $\delta \ll 1$ and $N\delta \gg 1$, Eq.~\ref{pRC} yields
\begin{equation}
p_{\mathrm{RC}} \approx \frac{\delta}{1-e^{-N \delta}}\approx \delta\,.
\end{equation}

Finally, $p_\mathrm{SC}$ denotes the fixation probability of a single C mutant in a population of S individuals, without antimicrobial. In this case, $f_\mathrm{S}=f_\mathrm{C}=1$, so we are in the neutral case, and Eq.~\ref{proba_A_neutral} yields $p_\mathrm{SC}=1/N$. 

\paragraph*{Fixation times.} First, $\tau_\mathrm{R}^\mathrm{d}$ denotes the average time it takes for the lineage of a single R mutant to disappear in the absence of antimicrobial. Hence, it is equal to the fixation time of the S type in a population that initially contains $N-1$ individuals of type S and 1 individual of type R. Considering S as type A and R as type B, we have $\gamma=f_\mathrm{S}/f_\mathrm{R}=1/(1-\delta)$ without antimicrobial, and our initial condition is $j_0=N-1$, so $\tau_\mathrm{R}^\mathrm{d}$ is equal to $t_{N-1}^\mathrm{S} /N$ (see Eq. \ref{t_N-1_A_select}). Recall that $t_{N-1}^\mathrm{S}$ needs to be divided by the population size $N$ because we expressed it in numbers of steps of the Moran process, while $\tau_\mathrm{R}^\mathrm{d}$ has to be expressed in numbers of generations. While the general formula Eq.~\ref{t_N-1_A_select} is rather complex, in the neutral case $\delta=0$, it reduces to the much simpler expression in Eq. \ref{t_N-1_A_neutral}, which yields $\tau_\mathrm{R}^\mathrm{d}\approx\log N$ for $N\gg 1$. For $\delta>0$, $\tau_\mathrm{R}^\mathrm{d}$ is shorter than in the neutral case, because the R mutants are out-competed by S individuals. Note that a good approximation to the exact formula in Eq. \ref{t_N-1_A_select} can be obtained within the diffusion approach~\cite{Ewens79} (see the Fokker-Planck equation below).

Second, $\tau^\mathrm{f}_\mathrm{R}$ denotes the average time needed for the R mutants take over with antimicrobial, starting from one R mutant and $N-1$ S individuals. Considering S as type A and R as type B, we have $\gamma=f_\mathrm{S}/f_\mathrm{R}=0$ with antimicrobial, and our initial condition is $j_0=N-1$. Then $\tau_\mathrm{R}^\mathrm{f}$ is equal to $t_{N-1}^\mathrm{R} /N$ (see Eq. \ref{t_1_R_select}), with $\gamma=0$. Using Eq. \ref{t_1_R_select}, we obtain
\begin{equation}
\displaystyle \tau_\mathrm{R}^\mathrm{f}=\sum_{i=1}^{N-1}\frac{1}{i}\,,
\end{equation}
which entails $\tau^\mathrm{f}_\mathrm{R}\approx\log N$ for $N\gg 1$.

Finally, $\tau^\mathrm{f}_\mathrm{C}$ denotes the average time needed for the C mutants to take over, starting from one C mutant and $N-1$ R individuals. Considering R as type A and C as type B, we have $\gamma=f_\mathrm{R}/f_\mathrm{C}=1-\delta$, independent whether antimicrobial is present or absent, and our initial condition is $j_0=N-1$. Hence, $\tau_\mathrm{C}^\mathrm{f}$ is given by $t_{N-1}^\mathrm{C} /N$ (see Eq. \ref{t_1_R_select}). In the neutral case $\delta=0$, $t_{N-1}^C$ reduces to Eq.~\ref{t_N-1_B_neutral}, and thus $\tau^\mathrm{f}_\mathrm{C}\approx N$ for $N\gg 1$. For $\delta>0$, it is shorter, as selection favors the fixation of C, and again a good approximation to the exact formula in Eq. \ref{t_1_R_select} can be obtained within the diffusion approach~\cite{Ewens79} (see the Fokker-Planck equation below). 

\section{Large populations: deterministic limit}
\label{SI_Det}

If stochastic effects are neglected, the dynamics of a microbial population can be described by coupled differential equations on the numbers of individuals of each  genotype~\cite{Ewens79}. This deterministic approach is appropriate if the number $N$ of competing microorganisms satisfies $N\mu_1\gg 1$~\cite{Rouzine01}. Here, we derive and study the deterministic limit of the complete stochastic model studied in the main text.

\subsection{From the stochastic model to the deterministic limit}

Here, we present a full derivation of the deterministic limit of the stochastic model based on the Moran process (see above). This derivation closely follows those of Refs.~\cite{Traulsen05,Traulsen09} and is presented here for the sake of pedagogy and completeness. Starting from the Master equation of our stochastic model, we obtain a Fokker-Planck equation, corresponding to the diffusion approximation~\cite{Ewens79}, and then a deterministic differential equation, in the limits of increasingly large population sizes.

Let us first recall the Master equation corresponding to the Moran process, where $j$ denotes the number of A individuals and $N-j$ the number of B individuals, as above:
\begin{equation}
\frac{dP_j(t)}{dt}=P_{j-1}(t)\,\Pi_{j-1\rightarrow j}+P_{j+1}(t)\,\Pi_{j+1\rightarrow j}-P_j(t)\left(\Pi_{j\rightarrow j-1}+\Pi_{j\rightarrow j+1}\right)\,.
\label{me}
\end{equation}
The notations in Eq.~\ref{me} are the same as in the previous section, and time is expressed in number of steps of the Moran process. Let us now introduce the reduced variables $x=j/N$, $\tau =t/N$, as well as $\rho (x,\tau )=N P_j(t)$. Then, since one step of the Moran process occurs each time unit, Eq. \ref{me} can be rewritten as:
\begin{align}
\rho(x,\tau+1/N)-\rho(x,\tau)=&\,\rho(x-1/N,\tau)\,\Pi^+(x-1/N)+\rho(x+1/N,\tau)\,\Pi^-(x+1/N) \nonumber \\
&-\rho(x,\tau)\left(\Pi^-(x)+\Pi^+(x)\right)\,,
\end{align}
with 
\begin{equation}
\Pi^-(x)=\Pi_{j\rightarrow j-1}=\frac{f_B x(1-x)}{f_A\,x+f_B\,(1-x)} \,\textrm{ and }\, \Pi^+(x)=\Pi_{j\rightarrow j+1}=\frac{f_A x(1-x)}{f_A\,x+f_B\,(1-x)}\,.
\end{equation}

\paragraph*{Diffusion approximation.} For $N \gg 1$, considering that jumps are small at each step of the Moran process, i.e. $1/N\ll x$ and $1/N\ll \tau$, the probability density $\rho(x,\tau)$ and the transition probabilities $\Pi^ \pm(x)$ can be expanded in a Taylor series around $x$ and $\tau$. This expansion, known as a Kramers-Moyal expansion~\cite{Gardiner}, yields, to first order in $1/N$: 
\begin{equation}
\frac{\partial \rho(x,\tau)}{\partial \tau}=-\frac{\partial}{\partial x}\left[\rho(x,\tau)a(x)\right]+\frac{1}{2}\frac{\partial ^2}{\partial x^2}\left[\rho(x,\tau)b^2(x)\right]
\label{fp}
\end{equation}
with 
\begin{equation}
a(x)=\Pi^+(x)-\Pi^-(x) \,\textrm{ and }\, b^2(x)=\frac{\Pi^+ (x) + \Pi^- (x) }{N}\,.
\end{equation}
Eq.~\ref{fp} is known as a diffusion equation, or a Fokker-Planck equation, or a Kolmogorov forward equation~\cite{Gardiner}, and $a(x)$ corresponds to the selection term (known as the drift term in physics), while $b^2(x)$ corresponds to the genetic drift term (known as the diffusion term in physics). 

\paragraph*{Deterministic limit.} In the limit $N\rightarrow\infty$, retaining only the zeroth-order terms in $1/N$, Eq. \ref{fp} reduces to:
\begin{equation}
\frac{\partial \rho (x,\tau)}{\partial \tau}=-\frac{\partial}{\partial x}\left[\rho(x,\tau)a(x)\right]\,.
\end{equation}
Let us focus on the average value of $x$, denoted by $\langle x \rangle$. Using Eq. \ref{fp} yields
\begin{align}
\frac{d \langle x \rangle}{d \tau}&=\int_0^1 \frac{\partial \rho (x,\tau)}{\partial \tau}\,x\,dx=-\int_0^1 \frac{\partial }{\partial x}\left[\rho (x,\tau)\,a(x)\right]\,dx\\
&=-\left[x\, \rho(x,\tau) \,a(x)\right]_0^1+\int_0^1 \rho(x,\tau)\,a(x)\,dx \label{IBP} \\
&=\langle a(x) \rangle
\end{align}
The first term of right hand side of Eq. \ref{IBP} vanishes because $a(0)=a(1)=0$. In the limit $N\rightarrow\infty$, the distribution of $x$ is very peaked around its mean, so $\langle x \rangle \approx x$ and $\langle a(x) \rangle \approx a(x)$, yielding:
\begin{equation}
\frac{dx}{d\tau}=x(1-x)\frac{\Delta f}{\bar{f}}\,,
\label{Modified_re}
\end{equation}
where $\Delta f=f_A-f_B$ denotes the difference of the fitnesses of the two types, while $\bar{f}=f_A\,x+f_B\,(1-x)$ is the average fitness in the population. Eq. \ref{Modified_re} is an ordinary differential equation known as the adjusted replicator equation~\cite{Traulsen05}. Recall that $\tau$ corresponds to the number $t$ of steps of the Moran process divided by the total number $N$ of individuals in the population. Hence, $\tau$ is the time in numbers of generations used in the main text, and Eq.~\ref{Modified_re} is the proper deterministic limit for our stochastic process. 

Note that in the framework of the Moran process, fitnesses are only relative. If one wanted to account for absolute fitness effects, so that a whole population reproduces faster if its average fitness is higher, one would need to include an additional rescaling of time $\tau' = \tau/ \bar{f}$. Note that if $\bar{f}$ is constant, this rescaling yields a standard replicator equation:
\begin{equation}
\frac{dx}{d\tau'}=x(1-x)\Delta f\,.
\label{replicator}
\end{equation}

\subsection{Deterministic description of the evolution of antimicrobial resistance}

\paragraph*{System of ordinary differential equations.}

Let us now come back to our model of the evolution of antimicrobial resistance, with three types of microorganisms (see Fig.~\ref{Fig1}A). In the limit of large populations, the  complete  stochastic model described in the main text will converge to a deterministic system of ordinary differential equations, as demonstrated above. Generalizing Eq.~\ref{replicator}, by considering three types of individuals and taking into account mutations, yields a system of replicator-mutator equations \cite{Traulsen09}:

\begin{equation}
\begin{cases}
\displaystyle
\dot{s}=f_\mathrm{S}(1-\mu_1)s-\overline{f}s \\
\displaystyle
\dot{r}=f_\mathrm{R}(1-\mu_2)r+ f_\mathrm{S} \, \mu_1 \, s -\overline{f} \, r \\
\displaystyle
s+r+c=1 \mbox{ },
\end{cases}
\label{pop_frac_syst}
\end{equation}
where $s$, $r$ and $c$ are the population fractions of S (sensitive), R (resistant) and C (resistant-compensated) microorganisms, respectively, while $f_\mathrm{S}$, $f_\mathrm{R}$ and $f_\mathrm{C}$ denote their fitnesses, $\overline{f}=f_\mathrm{S} \, s + f_\mathrm{R} \, r +f_\mathrm{C} \, c$ denotes the average fitness in the population, and dots denote time derivatives. To illustrate that Eq.~\ref{pop_frac_syst} generalizes Eq.~\ref{replicator}, consider the case where $c=0$ and $\mu_1=0$: the first equation of Eq.~\ref{replicator} then yields $\dot{s}=f_\mathrm{S} s-[f_\mathrm{S} s+f_\mathrm{R} (1-s)]s=s(1-s)(f_\mathrm{S}-f_\mathrm{R})$. As demonstrated above, the deterministic limit of our stochastic model yields adjusted replicator equations (see Eq.~\ref{Modified_re}). For the sake of simplicity, the present analytical discussion focuses on standard replicator equations (see Eq.~\ref{replicator}). 

The system of equations Eq.~\ref{pop_frac_syst} only concerns population fractions, and constitutes the large-population limit $N\rightarrow\infty$ of our stochastic model at constant $N$. It is mathematically convenient to note that the same equations are obtained in the case of a population in which microorganisms have an exponential growth. This model, which enables us to recover the system \ref{pop_frac_syst}, is governed by the following system of linear differential equations:
\begin{equation}
\begin{cases}
\displaystyle
\dot{N}_\mathrm{S}=f_\mathrm{S}(1-\mu_1)N_\mathrm{S} \\
\displaystyle
\dot{N}_\mathrm{R}=f_\mathrm{R}(1-\mu_2)N_\mathrm{R}+ f_\mathrm{S} \, \mu_1 \, N_\mathrm{S} \\
\displaystyle
\dot{N}_\mathrm{C}=f_\mathrm{C} \, N_\mathrm{C}+ f_\mathrm{R} \, \mu_2 \, N_\mathrm{R}\mbox{ },
\end{cases}
\label{deter_syst}
\end{equation}
where $N_\mathrm{S}$, $N_\mathrm{R}$ and $N_\mathrm{C}$ are the numbers of sensitive, resistant and resistant-compensated microorganisms, respectively. It is straightforward to show that the population fractions obtained from this exponential growth model satisfy Eq.~\ref{pop_frac_syst}: hence, this simple deterministic model allows one to understand the evolution of large microbial populations described by the Moran model (even though the total population is constant in the Moran model).

\paragraph*{Analytical resolution.}

Being linear, the system in Eq.~\ref{deter_syst} is straightforward to solve analytically:
\begin{equation}
\begin{pmatrix}
S \\ R \\ C
\end{pmatrix}
=
\begin{pmatrix}
0 & 0 & 1 \\
0 & 1 & \frac{f_\mathrm{S} \, \mu_1}{f_\mathrm{S}(1-\mu_1)-f_\mathrm{R}(1-\mu_2)} \\
1 & \frac{f_\mathrm{R} \, \mu_2}{f_\mathrm{R}(1-\mu_2)-f_\mathrm{C}} & \frac{f_\mathrm{S} \, \mu_1 \, f_\mathrm{R} \, \mu_2}{\left(f_\mathrm{S}(1-\mu_1)-f_\mathrm{R}(1-\mu_2)\right)\left(f_\mathrm{S}(1-\mu_1)-f_\mathrm{C}\right)}
\end{pmatrix}
\begin{pmatrix}
\beta_1 \, e^{f_\mathrm{C} \, t} \\ \beta_2 \, e^{f_\mathrm{R} (1-\mu_2)t} \\ \beta_3 \, e^{f_\mathrm{S} (1-\mu_1)t}
\end{pmatrix}
\label{gen_sol}
\end{equation}
where $\beta_1$, $\beta_2$ and $\beta_3$ can be expressed from the initial conditions $S(0)$, $R(0)$ and $C(0)$. The fractions $s$, $r$ and $c$ can then be obtained from this solution, e.g. through $s=S/(S+R+C)$. 

\paragraph*{Limiting regimes and characteristic timescales.} As in the main text, we are going to focus on the case where the population initially only comprises sensitive microorganisms, i.e. $s(0)=1$. In the case of periodic alternations of absence and presence of antimicrobial, a small fraction of R microorganisms will appear within the first half-period without antimicrobial. The subsequent evolution of the population composition can be separated into three successive regimes. In the first one, it suffices to consider S and R microorganisms, as the fraction of C is negligible, because the appearance of C requires an additional mutation. The second regime is more complex, and involves all three types of microorganisms, as the growth of C microorganisms makes the fractions of S and R microorganisms decrease. Then, provided that antimicrobial has been present for a sufficient time, the fraction of S microorganisms becomes negligible, because they cannot divide with antimicrobial. Hence, the third regime only involves R and C microorganisms, and does not depend on the presence or absence of antimicrobial, because the fitnesses of R and C are unaffected. Here, we determine analytically the main timescales involved in these first and third regimes. 

\paragraph*{First regime: S vs. R.}
Let us consider the first regime where there are almost only S and R microorganisms. We are interested in the population fractions $s(t)$ and $r(t)$, with $s(t)+r(t) \approx 1$. Eq.~\ref{pop_frac_syst} then gives:
\begin{equation}
\dot{s}=s\left(\Delta f_1 -s\, \Delta f_2\right) \,,
\label{ed_1}
\end{equation}
where we have defined $\Delta f_1=f_\mathrm{S}(1-\mu_1)-f_\mathrm{R}$ and $\Delta f_2=f_\mathrm{S}-f_\mathrm{R}$. Note that we expect $\Delta f_1 \approx \Delta f_2$, since biologically relevant values generally satisfy $\mu_1 \ll 1$ and $\mu_1 \ll \delta$. The solution of Eq.~\ref{ed_1} reads
\begin{equation}
s(t)=\frac{s_0\,e^{\Delta f_1 t}}{1-s_0\frac{\Delta f_2}{\Delta f_1}+s_0\frac{\Delta f_2}{\Delta f_1}\,e^{\Delta f_1 t}}\mbox{ },
\label{s_1}
\end{equation}
where $s_0$ is the fraction of S microorganisms at the beginning of the first regime (taken as $t=0$ here). In the presence of antimicrobial ($f_\mathrm{S}=0$), the previous expression can be simplified, using $\Delta f_1=\Delta f_2=-(1-\delta)$. This allows us to identify the characteristic time $\tau_1$ of the decay of $s$, as R microorganisms take over:

\begin{equation}
\tau_1=\frac{-1}{\Delta f_1}=\frac{1}{1-\delta} \mbox{ }.
\end{equation}

The duration $t_1$ of the first regime in the presence of antimicrobial is governed by $\tau_1$. More precisely, Eq. \ref{s_1} yields:

\begin{equation}
t_1=\frac{1}{1-\delta}\log\left(\frac{s_0\,(1-s_1)}{s_1\,(1-s_0)}\right)\mbox{ },
\label{t1}
\end{equation}

where $s_1$ is the fraction of S microorganisms at the end of the first regime, at which point the fraction of C microorganisms  is no longer negligible.

\paragraph*{Third regime: R vs. C.}

Let us now turn to the third regime, assuming that antimicrobial has been present  for a long enough time  to allow S microorganisms to become a small minority. 
Eq.~\ref{pop_frac_syst} then gives:
\begin{equation}
\dot{r}=r\left(\Delta f_3 -r\, \Delta f_4\right) \,,
\label{ed_2}
\end{equation}
with $\Delta f_3=f_\mathrm{R}(1-\mu_2)-f_\mathrm{C}=-\delta(1-\mu_2)-\mu_2$ and $\Delta f_4=f_\mathrm{R}-f_\mathrm{C}=-\delta$, independently of whether antimicrobial is present or not. Again, we  generally  expect $\Delta f_3 \approx \Delta f_4$. The solution of Eq.~\ref{ed_2} reads
\begin{equation}
r(t)=\frac{r_2\left(1-\mu_2+\mu_2/\delta\right)e^{-\left(\delta(1-\mu_2)+\mu_2\right)t}}{1-\mu_2+\mu_2/\delta-r_2+r_2\,e^{-\left(\delta(1-\mu_2)+\mu_2\right)t}}\underset{\mu_2 \ll \delta}{\underset{\mu_2 \ll 1}{\approx}}\frac{r_2\,e^{-\delta t}}{1-r_2+r_2\,e^{- \delta t}}\mbox{ },
\label{r_a}
\end{equation}
where $r_2$ is the fraction of R microorganisms at the beginning of the third regime (taken as $t=0$ here). Hence, the characteristic time $\tau_3$ of the decay of $r$ reads:
\begin{equation}
\tau_3=\frac{1}{\mu_2+\delta(1-\mu_2)}\underset{\mu_2 \ll \delta}{\underset{\mu_2 \ll 1}{\approx}} \frac{1}{\delta}\mbox{ }.
\end{equation}
The duration $t_3$ of the third regime in the presence of antimicrobial is governed by $\tau_3$. More precisely, Eq. \ref{r_a} yields:
\begin{equation}
t_3\underset{\mu_2\ll \delta}{\underset{\mu_2 \ll 1}{\approx}}\frac{1}{\delta}\log\left(\frac{r_2\,(1-r_3)}{r_3\,(1-r_2)}\right)\mbox{ }.
\label{t3}
\end{equation}
where $r_3$ is the fraction of R microorganisms at the end of this regime, when C has become dominant in the population.

Note that the timescales obtained here are governed by selection (through the relevant fitness differences $\delta$ and $1-\delta$). This stands in contrast with the results from our stochastic model (see main text) where mutation rates are crucial, especially through the waiting time before resistant mutants appear. In the deterministic description considered here, small fractions of resistant mutants appear right away, so this consideration is irrelevant. However, mutation rates come into play in the durations of the different regimes within the deterministic model, through the fractions of each type of microorganisms at the beginning and at the end of each regime, but with a weak logarithmic dependence (see Eqs.~\ref{t1}-\ref{t3}).

\subsection{Comparison of stochastic and deterministic results}

As in the main text, we now focus on the impact of a periodic presence of antimicrobial on the time it takes for a population to fully evolve resistance. For large microbial populations satisfying $N\gg1/\mu_1$, we wish to check that the system of differential equations in Eq.~\ref{pop_frac_syst} recovers the results obtained with our stochastic model. To this end, we solve the system in Eq.~\ref{pop_frac_syst} numerically in the case of a periodic presence of antimicrobial. Note that complete fixation of a genotype does not happen in the deterministic model. Conversely, in the stochastic model, for a population of size $N$, the fixation of C corresponds to the discrete Moran step where the fraction $c$ jumps from $1-1/N$ to $1$. Hence, for our comparison between the deterministic results and the stochastic ones obtained for $N$ microorganisms, we consider that C effectively fixes in the deterministic model when the fraction $c$ reaches $1-1/N$. In addition, for exactness, we use a numerical resolution of the system in Eq.~\ref{pop_frac_syst} where time is rescaled through $t \rightarrow t / \bar{f}$. Indeed, the proper deterministic limit of our stochastic model corresponds to modified replicator equations, such as Eq. \ref{Modified_re} (see above). 

\begin{figure}[h!]
\begin{center}
   \includegraphics[width=0.95\textwidth]{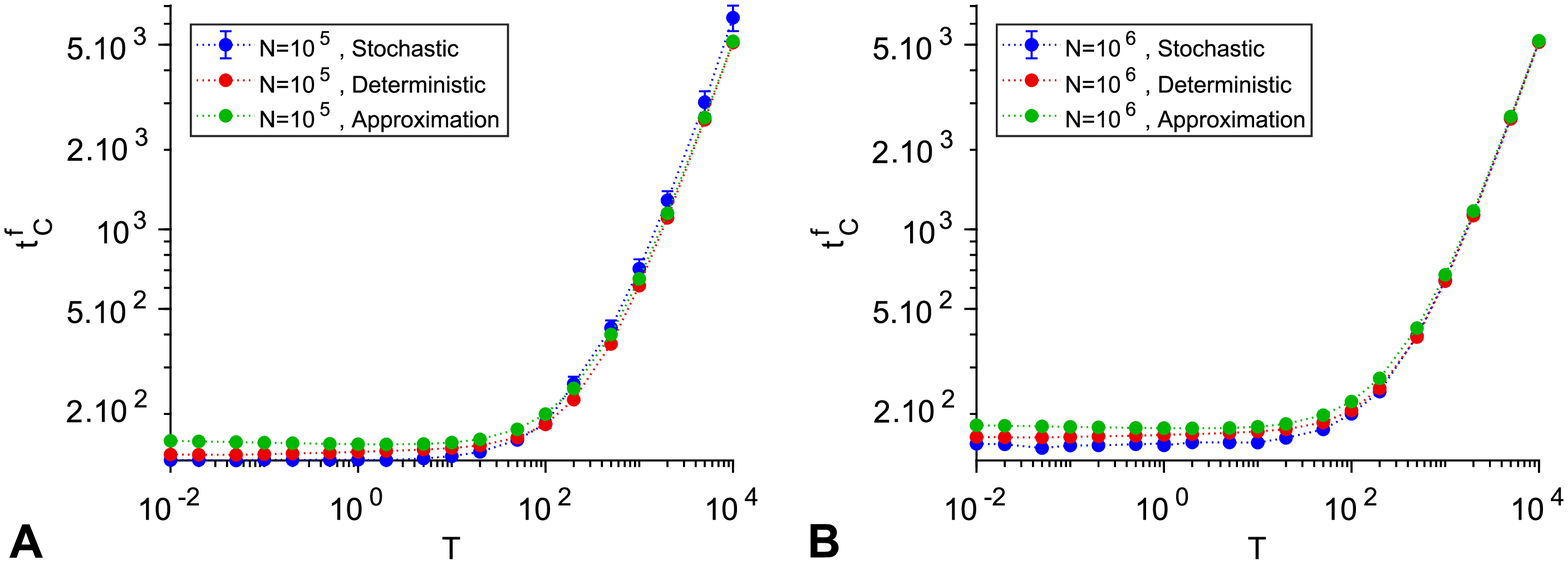}
   \quad
   \caption{\textbf{Large populations: stochastic model vs. deterministic model.} The total time $t_\mathrm{C}^\mathrm{f}$ of full resistance evolution is plotted versus the period $T$ of alternations of absence and presence of antimicrobial, in the case of symmetric alternations. Results from simulations of the stochastic model (see Fig.~\ref{varyT_and_varyN}A), numerical resolution of the deterministic model, and an analytical approximation of the deterministic solution (Eqs. \ref{t_C_f_deter}-\ref{rT/2}), are represented for $N=10^5$ (A) and $N=10^6$ (B). Parameter values: $\mu_1=10^{-5}$, $\mu_2=10^{-3}$, and $\delta=0.1$. }
   \label{det_vs_stoch}
\end{center}
\end{figure}

Fig.~\ref{det_vs_stoch} shows that the deterministic model yields results very close to those obtained through the stochastic model, in the case of large population sizes $N\geq 1/\mu_1$. We recover the regimes described in the main text, with a plateau for short periods, and a linear dependence on $T$ for larger ones. Moreover, the relative error made by using the deterministic model instead of the stochastic one is less than $\sim\!20\%$ (resp. $\sim\!10\%$) for all data points with $N=10^5$ (resp. $N=10^6$) in Fig.~\ref{det_vs_stoch}.

Let us now present an analytical approximation for $t_\mathrm{C}^\mathrm{f}$, based on the different timescales computed previously. As the population is initially only composed of S microorganisms, they will remain dominant during the first half-period without antimicrobial, since they are fitter than R mutants (and we assume that $T/2$ is not large enough to extend to the point where C starts being important, which would then correspond to the valley crossing case). Afterwards, R microorganisms start growing fast during the second half-period. Note that in the deterministic case, there is always a nonzero fraction of resistant microorganisms at the end of the first half-period without antimicrobial, contrary to the stochastic case studied in the main text. Hence, we compute the fraction $s_0=s(T/2)$ of S microorganisms at the end of the first half period, by using results for the above-described first regime without antimicrobials. This fraction $s_0=s(T/2)$ is then taken as the initial condition of the first regime with antimicrobial. Then, for simplicity, we assume that $s$ decays until it reaches $s_1\approx0.1$ (so $r_1\approx0.9$), while remaining in the first regime described above, in the presence of antimicrobial. We then assume the duration of the second regime is negligible, and consider that the third regime process starts right away, with a fraction $r_2\approx0.9$. As explained above, we consider that the third regime ends upon effective fixation of C, i.e. when $c$ reaches $1-1/N$, which implies $r_3=1/N$. Using Eqs.~\ref{t1} and~\ref{t3}, we obtain:
\begin{equation}
t_\mathrm{C}^\mathrm{f} \approx \frac{T}{2} + \frac{1}{1-\delta} \log\left( \frac{9\,s(T/2)}{1-s(T/2)} \right) + \frac{1}{\delta} \log \left( 9(N-1)  \right) \, ,
\label{t_C_f_deter}
\end{equation}
where $s(T/2)$ is obtained by using Eq.~\ref{s_1} in the absence of antimicrobial:
\begin{equation}
s(T/2)=\frac{e^{\left(\mu_2+\delta(1-\mu_2)-\mu_1\right)T/2}}{1-\frac{\mu_2 + \delta (1 - \mu_2)}{\mu_2 + \delta (1 - \mu_2)-\mu_1}\left(1-e^{\left(\mu_2+\delta(1-\mu_2)-\mu_1\right)T/2}\right)} \, .
\label{rT/2}
\end{equation}

Eqs. \ref{t_C_f_deter}-\ref{rT/2} yield good approximations of the analytical results obtained by numerical resolution of Eq.~\ref{pop_frac_syst}, as can be seen on Fig. \ref{det_vs_stoch}. More precisely, the relative error made by using this approximation instead of the full numerical resolution is less than $\sim\!13\%$ for all parameters in Fig.~\ref{det_vs_stoch}. 

For $T\gg2/\delta$, Eq.~\ref{rT/2} reduces to $s(T/2)\approx 1-\mu_1/[\mu_2+\delta(1-\mu_2)]\approx 1-\mu_1/\delta$, so only the first term in Eq.~\ref{t_C_f_deter} then depends on $T$. Hence, this term becomes dominant for large $T$, yielding $t_\mathrm{C}^\mathrm{f} \approx T/2$ in this limit. This asymptotic behavior is again consistent with our predictions from the stochastic model (see main text). The horizontal purple solid line at large $T$ in Fig.~\ref{varyT_and_varyN}A, and the horizontal solid lines at large $N$ in Fig.~\ref{varyT_and_varyN}B, both correspond to $t_\mathrm{C}^\mathrm{f} \approx T/2$, showing excellent agreement with our stochastic simulations as well.

Conversely, for small periods, the first term of Eq.~\ref{t_C_f_deter} can be neglected, so the dependence on $T$ of $t_\mathrm{C}^\mathrm{f}$ is weaker (Eq.~\ref{rT/2} reduces to $s(T/2)\approx 1-\mu_1 T/2$ for $T\ll2/\delta$, so a weak logarithmic dependence on $T$ remains, due to the second term of Eq.~\ref{t_C_f_deter}). It is interesting to note that the third term of $t_\mathrm{C}^\mathrm{f}$ in Eq.~\ref{t_C_f_deter} also increases logarithmically with $N$. This stands in contrast with the case of smaller populations, where our stochastic study showed that $t_\mathrm{C}^\mathrm{f}$ essentially decreases linearly with $N$ (see main text). This change of behavior as $N$ increases can be seen on Fig.~\ref{varyT_and_varyN}A in the regime of small $T$ (in particular, for large $N$, the purple data points corresponding to $N=10^6$ are then slightly higher than the blue ones corresponding to $N=10^5$; see also Fig.~\ref{det_vs_stoch}, where the y-axis range and scale are the same on panels A and B).

\section{Comparison to spontaneous fitness valley crossing}
\label{SI_Valley}

\subsection{No antimicrobial: Crossing of a symmetric fitness valley}

Let us compare the alternation-driven evolution of resistance to what would happen in the absence of alternations of phases of absence and presence of antimicrobial. If a population composed only of S (sensitive) microorganisms is subjected to a continuous presence of antimicrobial, it will not evolve resistance, because divisions are blocked (see Fig.~\ref{Fig1}A). Conversely, a population of S microorganisms that is never subjected  to antimicrobial can spontaneously evolve resistance. In our model, this will eventually happen. This process is difficult and slow, because of the initial fitness cost of resistance: it requires crossing a fitness valley (see Fig.~\ref{Fig1}A). Fitness valley crossing has been studied in detail~\cite{Nowak02,Weinreich05,Weissman09,Weissman10,Bitbol14}, but usually in the case where the final mutant has a higher fitness than the initial organism. In the evolution of antimicrobial resistance, compensatory mutations generally yield microorganisms with antimicrobial-free fitnesses that are similar to, but not higher than those of sensitive microorganisms~\cite{Borman96,Schrag97,Andersson10}. Hence, we here extend the known results for fitness valley crossing by constant-size homogeneous asexual populations~\cite{Weissman09} to ``symmetric'' fitness valleys, where the final genotype has no selective advantage compared to the initial one. Briefly, the main difference with Ref.~\cite{Weissman09} is that the probability of establishment of the second mutant (C) in a population with a majority of non-mutants (S) is $1/N$ instead of being given by the selective advantage $s$ of the second mutant. This probability plays an important role in the tunneling case.

There are two different ways of crossing a fitness valley. In \textit{sequential fixation}, the first deleterious mutant fixes in the population, and then the second mutant fixes. In \textit{tunneling}~\cite{Nowak02}, the first mutant never fixes in the population, but a lineage of second mutants arises from a minority of first mutants, and fixes. For a given valley, characterized by $\delta$ (see Fig.~\ref{Fig1}A), population size $N$ determines which mechanism dominates. Sequential fixation requires the fixation of a deleterious mutant through genetic drift, and dominates for small $N$, when stochasticity is important. Tunneling dominates above a certain $N$~\cite{Weinreich05,Weissman09}. Let us study these two mechanisms in the regime of rare mutations $N\mu_1\ll 1$ where stochasticity is crucial.

In sequential fixation, the average time $\tau_\mathrm{SF}$ to cross a valley is the sum of those of each step involved~\cite{Weissman09}. Hence $\tau_\mathrm{SF}=1/(N\mu_1p_\mathrm{SR})+1/(N\mu_2p_\mathrm{RC})$, where $p_\mathrm{SR}$ (resp. $p_\mathrm{RC}$) is the fixation probability of a single R (resp. C) individual in a population of size $N$ where all other individuals are of type S (resp. R). Fixation probabilities are known in the Moran process (see Supplementary Material, Section~\ref{SI_Moran}). In particular, if $N\delta\ll 1$ then $p_\mathrm{SR}\approx p_\mathrm{RC}\approx1/N$ for our symmetric valley, so $\tau_\mathrm{SF}\approx1/\mu_1+1/\mu_2$ ($\approx1/\mu_1$ if $\mu_1\ll\mu_2$), while if $\delta\ll 1$ and $N\delta\gg 1$ then $p_\mathrm{SR}\approx\delta e^{-N\delta}$ and $p_\mathrm{RC}\approx\delta\gg p_\mathrm{SR}$, so $\tau_\mathrm{SF}\approx e^{N\delta}/(N\mu_1\delta)$.

In tunneling, the key timescale is that of the appearance of a successful first (R) mutant, i.e. a first mutant whose lineage will give rise to a second (C) mutant that will fix in the population~\cite{Weissman09}. Neglecting subsequent second mutation appearance and fixation times, the average tunneling time reads $\tau_\mathrm{T}\approx 1/(N\mu_1 p_1)$, where $p_1$ is the probability that a first mutant is successful~\cite{Weissman09}. Upon each division of a first mutant, the probability of giving rise to a second mutant that will fix is $p=\mu_2 p_\mathrm{SC}$, where $p_\mathrm{SC}$ is the fixation probability of a single C mutant in a population of S individuals. For our symmetric valley, $p_\mathrm{SC}=1/N$, so $p=\mu_2/N$. In the neutral case $\delta=0$, Ref.~\cite{Weissman09} demonstrated that the first-mutant lineages that survive for at least $\sim\!1/\sqrt{p}$ generations, and reach a size $\sim\!1/\sqrt{p}$, are very likely to be successful, and fully determine the rate at which successful first mutants are produced. Since the lineage of each new first mutant has a probability $\sim\! \sqrt{p}$ of surviving for at least $\sim\!1/\sqrt{p}$ generations~\cite{Weissman09}, the probability that a first mutant is successful is $p_1\sim\!\sqrt{p}\sim\!\sqrt{\mu_2/N}$. If $\delta>0$, a first mutant remains effectively neutral if its lineage size is smaller than $1/\delta$~\cite{Weissman09}. Hence, if $\delta<\sqrt{\mu_2/N}$, $p_1\sim\!\sqrt{\mu_2/N}$ still holds. (This requires $N\mu_2\gg 1$, otherwise the first mutant fixes before its lineage reaches a size $\sqrt{N/\mu_2}$.) Finally, if $\delta>\sqrt{\mu_2/N}$, the lineage of a first mutant will reach a size at most $\sim\!1/\delta$, with a probability $\sim\!\delta$ and a lifetime $\sim\!1/\delta$~\cite{Weissman09}, yielding $p_1\sim\! \mu_2/(N\delta)$. 

Given the substantial cost of resistance mutations ($\delta\sim 0.1$~\cite{Schrag97,Andersson10}) and the low compensatory mutation rates (in bacteria $\mu_2\sim 10^{-8}$~\cite{Andersson10}), let us henceforth focus on the case where $\delta>\sqrt{\mu_2/N}$ (which is appropriate for all $N\geq 1$ with the values mentioned). Then $\tau_\mathrm{T}\approx 1/(N\mu_1 p_1)\approx \delta/(\mu_1 \mu_2)$, and two extreme cases can be distinguished:

(A) $N\delta\ll 1$ (effectively neutral regime): Then, $\tau_\mathrm{SF}\approx1/\mu_1$ (for $\mu_1\ll\mu_2$) and $\tau_\mathrm{T}\approx \delta/(\mu_1 \mu_2)$. Given the orders of magnitude above, generally $\delta>\mu_2$ in resistance evolution. Hence, sequential fixation is fastest, and the valley crossing time $\tau_\mathrm{V}$ reads:
\begin{equation}
\tau_\mathrm{V}=\tau_\mathrm{SF}\approx\frac{1}{\mu_1}\,. \label{neutral_seq_fix}
\end{equation}

(B) $\delta\ll 1$ and $N\delta\gg 1$: Then, 
\begin{equation}
\tau_\mathrm{V}=\min\left(\tau_\mathrm{SF},\,\tau_\mathrm{T}\right)\approx\min\left(\frac{e^{N\delta}}{N\mu_1\delta},\,\frac{\delta}{\mu_1\mu_2}\right)\,. \label{deleterious_valley_crossing}
\end{equation}
The transition from sequential fixation to tunneling~\cite{Weissman09} occurs when $N\delta e^{-N\delta}=\mu_2/\delta$. 

We have focused on the rare mutation regime $N\mu_1\ll 1$. If mutations are more frequent, the first successful lineage of R mutants that appears may not be the one that eventually fixes, so the valley-crossing time becomes shorter~\cite{Weissman09}.

In Fig.~\ref{varyT_and_varyN}B, the black simulation data points were obtained without any antimicrobial. The population then evolves resistance by valley crossing. The black curves correspond to our analytical predictions in Eq.~\ref{neutral_seq_fix} for $N\ll 1/\delta$ and in Eq.~\ref{deleterious_valley_crossing} for  $N\gg 1/\delta$. In the latter regime, the transition from sequential fixation to tunneling occurs at $N\approx65$ for the parameters of Fig.~\ref{varyT_and_varyN}B. The agreement between simulation results and analytical predictions is excellent, with no adjustable parameter.

\subsection{Alternation-driven process vs. valley-crossing process}

Now that we have studied the spontaneous crossing of a symmetric fitness valley without any antimicrobial, let us come back to our periodic alternations of phases of absence and presence of antimicrobial. Resistance can then evolve by two distinct mechanisms, namely the alternation-driven process and the spontaneous valley-crossing process. It is important to compare the associated timescales, in order to assess which process will happen faster and dominate. This will shed light on the acceleration of resistance evolution by the alternations. For generality, we consider asymmetric alternations.

With alternations, spontaneous valley crossing can still happen, but new R lineages cannot appear with antimicrobial, because S individuals cannot divide (see Fig.~\ref{Fig1}A). Since the appearance of a successful R mutant is usually the longest step of valley crossing (see above), the average valley crossing time $\tau'_\mathrm{V}$ with alternations will be longer by a factor $T/T_1$ than that without antimicrobial ($\tau_\mathrm{V}$), if more than one antimicrobial-free phase is needed to cross the valley, i.e. if $T_1\ll\tau_\mathrm{V}$. Eqs.~\ref{neutral_seq_fix} and~\ref{deleterious_valley_crossing} then yield
\begin{align}
\tau'_\mathrm{V}&\approx\frac{T}{T_1\mu_1}&\textrm{ for }&N\delta\ll 1\,, \label{neutral_seq_fix_alt}\\
\tau'_\mathrm{V}&\approx\frac{T}{T_1}\min\left(\frac{e^{N\delta}}{N\mu_1\delta},\,\frac{\delta}{\mu_1\mu_2}\right)&\textrm{ for }&\delta\ll 1\textrm{ and }N\delta\gg 1\,. \label{deleterious_valley_crossing_alt}
\end{align}
Conversely, if $T_1\gg\tau_\mathrm{V}$, valley crossing generally happens within the first antimicrobial-free phase. Hence, the average valley crossing time $\tau_\mathrm{V}$ is given by Eqs.~\ref{neutral_seq_fix} and~\ref{deleterious_valley_crossing}. (Recall that the process is assumed to begin with an antimicrobial-free phase.)

We can now compare the timescales of the valley-crossing process to those of the alternation-driven process. For simplicity, let us assume that the dominant timescale in the latter process is the time $t_\mathrm{R}^\mathrm{a}$ it takes to first observe an R organism in the presence of antimicrobial, i.e. $t_\mathrm{C}^\mathrm{f}\approx t^\mathrm{a}_\mathrm{R}$ (see Eq.~\ref{TotTime}). This is the case in a large and relevant range of parameters, especially if $\mu_1\ll\mu_2$, as discussed above. Note also that the final step of fixation of the successful C lineage, which can become long in large populations (up to $\sim\!N$ in the neutral case, see Supplementary Material, Section~\ref{SI_Moran}), is the same in the alternation-driven process and in the valley-crossing process, so it does not enter the comparison. The expression of $t^\mathrm{a}_\mathrm{R}$ in Eq.~\ref{tR_asymm} should thus be compared to the valley crossing time. If $T_1\gg\tau_\mathrm{V}$, valley crossing happens before any alternation, and is thus the relevant process, with time $\tau_\mathrm{V}$ given by Eqs.~\ref{neutral_seq_fix} and~\ref{deleterious_valley_crossing}. Let us now conduct our comparison of $t^\mathrm{a}_\mathrm{R}$ and $\tau'_\mathrm{V}$ for $T_1\ll\tau_\mathrm{V}$, where Eqs.~\ref{neutral_seq_fix_alt} and~\ref{deleterious_valley_crossing_alt} hold.

(A) If $T_1\ll\tau_\mathrm{R}^\mathrm{d}$ (recall that $\tau_\mathrm{R}^\mathrm{d}$ is the average lifetime of an R lineage without antimicrobial, before it goes extinct): The alternation-driven process, with timescale $t^\mathrm{a}_\mathrm{R}=T/(N\mu_1 T_1)$ (see Eq.~\ref{tR_asymm}), dominates. Indeed, if $N\delta\ll 1$, $\tau'_\mathrm{V}$ is given by Eq.~\ref{neutral_seq_fix_alt}, so for all $N> 1$, $t^\mathrm{a}_\mathrm{R}<\tau'_\mathrm{V}$. And if $N\delta\gg 1$  and $\delta\ll 1$, Eq.~\ref{deleterious_valley_crossing_alt} yields $t^\mathrm{a}_\mathrm{R}/\tau'_\mathrm{V}\approx \delta e^{-N\delta}\ll  1$ in the sequential fixation regime, and $t^\mathrm{a}_\mathrm{R}/\tau'_\mathrm{V}\approx \mu_2/(N\delta)\ll  1$ in the tunneling regime. Hence, if $T_1\ll\tau_\mathrm{R}^\mathrm{d}$, the alternation-driven process dominates. Thus, alternations of absence and presence of antimicrobial strongly accelerate resistance evolution. For instance, in Fig.~\ref{varyT_and_varyN}A, for $N=100$ and $T/2\ll\tau_\mathrm{R}^\mathrm{d}$, the alternation-driven process takes $t^\mathrm{a}_\mathrm{R}=2/(N\mu_1)=2\times 10^3$ generations, while valley crossing takes $\tau_\mathrm{V}=\delta/(\mu_1\mu_2)=10^7$ generations without antimicrobial: alternations yield a speedup of 4 orders of magnitude. The speedup is even stronger for larger populations. Conversely, for $T_1\ll T_2$, while the alternation-driven process is shorter than the valley-crossing process in the presence of alternations, it can nevertheless be longer than the valley-crossing process in the absence of antimicrobial. In this case, the drug actually slows down the evolution of resistance. When $T_1\ll T_2$ and $T_1\ll\tau_\mathrm{R}^\mathrm{d}$, in the tunneling regime, provided that $1/N\ll\delta\ll2\sqrt{\mu_2/N}$, valley crossing takes $\delta/(\mu_1\mu_2)$ in the absence of antimicrobial (see Eq.~\ref{deleterious_valley_crossing}), and $T_2\delta/(T_1\mu_1\mu_2)$ in the presence of alternations satisfying $T_1\ll T_2$ (see Eq.~\ref{deleterious_valley_crossing_alt}). Meanwhile, the switch-driven process takes $T_2/(T_1N\mu_1)$ (see above). Hence, if $T_2\gg T_1N\delta/\mu_2$, the alternation-driven process dominates, but it is slower than the valley-crossing process in the absence of antimicrobial: the drug then slows down resistance evolution. This effect can be seen on Fig.~\ref{varyT1_and_varyT2} for $T_1\ll T_2$ and $T_1\ll\tau_\mathrm{R}^\mathrm{d}$.

(B) If $T_1\gg\tau_\mathrm{R}^\mathrm{d}$: Then $t^\mathrm{a}_\mathrm{R}=T/(N\mu_1 \tau_\mathrm{R}^\mathrm{d})$ (see Eq.~\ref{tR_asymm}). If $N\delta\ll 1$, valley crossing by sequential fixation is the dominant process. Indeed, Eq.~\ref{neutral_seq_fix_alt} yields $t^\mathrm{a}_\mathrm{R}/\tau'_\mathrm{V}\approx T_1/\tau_\mathrm{R}^\mathrm{d}\gg 1$. If $N\delta\gg 1$  and $\delta\ll 1$, Eq.~\ref{deleterious_valley_crossing_alt} yields $t^\mathrm{a}_\mathrm{R}/\tau'_\mathrm{V}\approx \delta e^{-N\delta}T_1/\tau_\mathrm{R}^\mathrm{d}$ in the sequential fixation regime, and $t^\mathrm{a}_\mathrm{R}/\tau'_\mathrm{V}\approx \mu_2 T_1/(N\delta \tau_\mathrm{R}^\mathrm{d})$ in the tunneling regime. A transition from the alternation-driven process to valley crossing occurs when these ratios reach $1$. Qualitatively, if $N$ is large enough and/or if $T_1$ is short enough, the alternation-driven process dominates.

For example, in Fig.~\ref{varyT1_and_varyT2}A, parameters are such that the dominant mechanism of valley crossing is tunneling, so $t^\mathrm{a}_\mathrm{R}/\tau'_\mathrm{V}$ reaches 1 for $T_1=N\delta \tau_\mathrm{R}^\mathrm{d}/\mu_2\approx 2.6\times10^5$ generations. This transition to the valley-crossing plateau is indeed observed for the curves with large enough $T_2$. (Recall that if $T_2\ll\tau_\mathrm{R}^\mathrm{f}$, extinction events occur when $T_1\gg\tau_\mathrm{R}^\mathrm{d}$, see Fig.~\ref{Fluct_and_ext}B.) The black horizontal lines in Figs.~\ref{varyT1_and_varyT2}A and~\ref{varyT1_and_varyT2}B correspond to our analytical prediction in Eq.~\ref{deleterious_valley_crossing_alt}, giving $\tau'_\textrm{V}\approx \delta/(\mu_1\mu_2)$ if $T_1\gg \max(T_2, \tau_\mathrm{R}^\mathrm{d})$. Similarly, in Fig.~\ref{varyT_and_varyN}A, horizontal solid lines at large $T$ correspond to the valley crossing times in Eqs.~\ref{neutral_seq_fix_alt} or~\ref{deleterious_valley_crossing_alt}, depending on $N$. In Fig.~\ref{varyT_and_varyN}B, in the regime of small $N$ and large $T$, resistance evolution is achieved by tunneling-type valley crossing, yielding a plateau in the neutral regime $N \ll 1/\delta$ (see Eq.~\ref{neutral_seq_fix_alt}, plotted as a horizontal purple line) and an exponential increase for intermediate $N$ (see Eq.~\ref{deleterious_valley_crossing_alt}). For larger $N$, we observe a $T$-dependent transition to the alternation-driven process, which can be fully understood using the ratio $t^\mathrm{a}_\mathrm{R}/\tau'_\mathrm{V}$ (see above).

 \section{Detailed analysis of asymmetric alternations}
\label{SI_Asym}

\subsection{Particular regimes}

\begin{figure}[h!]
\centering
 \includegraphics[width=1.0\linewidth]{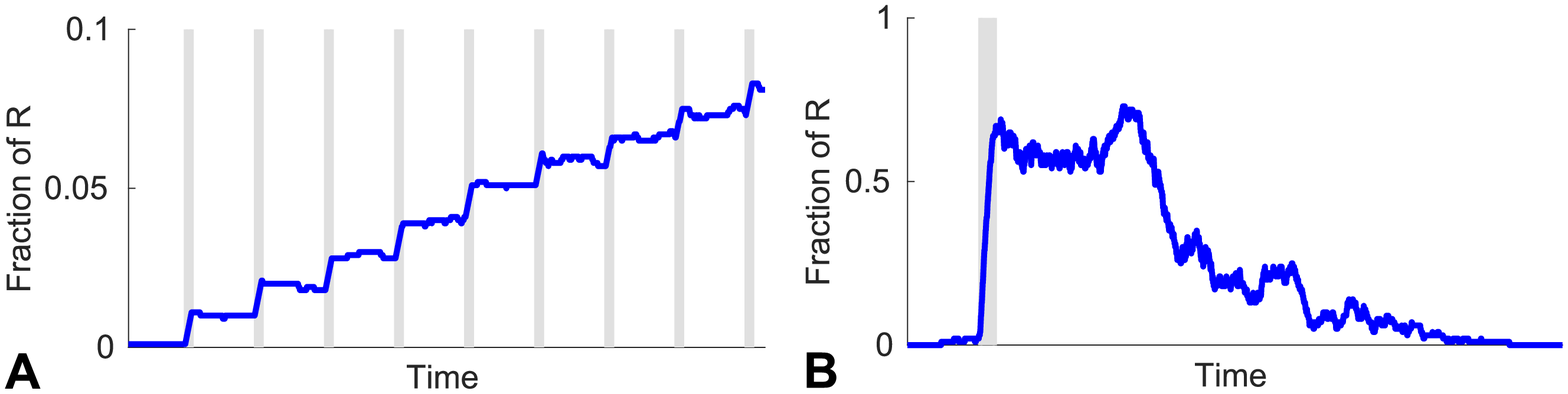}
 \quad
\caption{\textbf{Particular regimes}. The number of R individuals in the population is plotted versus time under alternations of phases without (white) and with antimicrobial (gray). Data extracted from simulation runs. (A) $T_2\ll\tau^\mathrm{f}_\mathrm{R}$ and $T_1\ll\tau^\mathrm{d}_\mathrm{R}$: the R lineage drifts for multiple periods. Parameters: $N=10^3$, $T_1=10^{-1}$, $T_2=10^{-2}$ . (B) $T_2\ll\tau^\mathrm{f}_\mathrm{R}$ and $T_1\gg\tau^\mathrm{d}_\mathrm{R}$: the R lineage goes extinct. Parameters: $N=10^2$, $T_1=10^{2}$, $T_2=1$. In both (A) and (B), $\mu_1=10^{-5}$, $\mu_2=10^{-3}$ and $\delta=0.1$.}
\label{Fluct_and_ext}
\end{figure}

Here, we examine whether R mutants will fix during a single phase with antimicrobial, of duration $T_2$. The fixation time of the lineage of an R mutant in the presence of antimicrobial is $\tau^\mathrm{f}_\mathrm{R}\approx\log N$ for $N\gg 1$~\cite{Ewens79} (see above). If $T_2\gg\tau^\mathrm{f}_\mathrm{R}$, fixation will happen within $T_2$. In the opposite case, the fixation of R is not likely to occur within a single phase with antimicrobial. Two situations exist in this case (see Fig.~\ref{Fluct_and_ext}). 

(A) If $T_2\ll\tau^\mathrm{f}_\mathrm{R}$ and $T_1\ll\tau^\mathrm{d}_\mathrm{R}$ (Fig.~\ref{Fluct_and_ext}A): The R lineage will drift for multiple periods, but its extinction is unlikely, as for symmetric alternations. This effect can induce a slight increase of the total time of resistance evolution, which is usually negligible.

(B) If $T_2\ll\tau^\mathrm{f}_\mathrm{R}$ and $T_1\gg\tau^\mathrm{d}_\mathrm{R}$ (Fig.~\ref{Fluct_and_ext}B): The R lineage is likely to go extinct even after it has started growing in the presence of antimicrobial. This typically implies $T_1\gg T_2$, since $\tau^\mathrm{f}_\mathrm{R}\approx\log N$ and $\tau^\mathrm{d}_\mathrm{R}\lesssim\log N$ for $N\gg 1$ (see above). Hence, this case is specific to (very) asymmetric alternations. Spontaneous valley crossing then becomes the fastest process of resistance evolution (see Supplementary Material, Section~\ref{SI_Valley}).

\subsection{Varying $T_2$ at fixed $T_1$}

In the main text, we present a detailed analysis of what happens when $T_1$ is varied at fixed $T_2$ (see Fig.~\ref{varyT1_and_varyT2}A). Here, we present a similar analysis if $T_2$ is varied at fixed $T_1$. Fig.~\ref{varyT1_and_varyT2}B shows the corresponding simulation results, together with our analytical predictions from Eqs.~\ref{TotTime} and~\ref{tR_asymm}. In particular, a minimum is observed in Fig. \ref{varyT1_and_varyT2}B when varying $T_2$ for $T_1\gg\tau_\mathrm{R}^\mathrm{d}$:
\begin{itemize}
\item When $T_2\ll\tau_\mathrm{R}^\mathrm{d}\ll T_1$, valley crossing dominates. 
\item When  $\tau_\mathrm{R}^\mathrm{d}\ll T_2\ll T_1$, Eq.~\ref{tR_asymm} gives $t^\mathrm{a}_\mathrm{R}= T/(N\mu_1\tau_\mathrm{R}^\mathrm{d})\approx T_1/(N\mu_1\tau^\mathrm{d}_\mathrm{R})$, which is independent from $T_2$. \item As $T_2$ is further increased, $t^\mathrm{a}_\mathrm{R}= T/(N\mu_1\tau_\mathrm{R}^\mathrm{d})$ increases, becoming proportional to $T_2$ when $T_2\gg T_1$. 
\end{itemize}
Hence, the minimum of $t^\mathrm{a}_\mathrm{R}$ is $T_1/(N\mu_1\tau^\mathrm{d}_\mathrm{R})$ and is attained for $\tau_\mathrm{R}^\mathrm{d}\ll T_2\ll T_1$. 
In the opposite case where $T_1\ll\tau_\mathrm{R}^\mathrm{d}$, Eq.~\ref{tR_asymm} still gives $t^\mathrm{a}_\mathrm{R}= T/(N\mu_1T_1)$. Thus, $t^\mathrm{a}_\mathrm{R}$ reaches a plateau $t^\mathrm{a}_\mathrm{R}=1/(N\mu_1)$ for $T_2\ll T_1\ll\tau_\mathrm{R}^\mathrm{d}$, which means that the first R mutant yields the full evolution of resistance (as seen above). Then, $t^\mathrm{a}_\mathrm{R}$ becomes proportional to $T_2$ for $T_2\gg T_1$. Note that valley crossing is always slower than the alternation-driven process when $T_1\ll\tau_\mathrm{R}^\mathrm{d}$ (see above), so no plateau is expected at large $T_2$ in this case.

\section{Robustness of the binary antimicrobial action model}
\label{SI_Robust}

Throughout our study, we have modeled the action of the antimicrobial in a binary way: below the MIC (``absence of antimicrobial''), growth is not affected, while above it (``presence of antimicrobial''), sensitive microorganisms cannot grow at all (see Model section in the main text). The relationship between antimicrobial concentration and microorganism fitness is termed the pharmacodynamics of the antimicrobial~\cite{Jacobs01,Regoes04}. Our binary approximation is motivated by the usual steepness  of pharmacodynamic curves around the MIC~\cite{Regoes04}. However, this steepness  is not infinite, and it is different for each antimicrobial. Here, we investigate the robustness of our binary model. 

If one goes beyond the binary model and accounts for the smoothness of the pharmacodynamic curve, one additional factor enters the determination of the time dependence of fitness. It is the time dependence of the antimicrobial concentration, typically in a treated patient, which is known as pharmacokinetics~\cite{Jacobs01,Regoes04}. In fact, the time dependence of the fitness of sensitive microorganisms will be determined by a combination of pharmacodynamics and pharmacokinetics. Experimental pharmacodynamic curves are well-fitted by Hill functions, and pharmacokinetic curves are often modeled by exponential decays of drug concentration after intake~\cite{Regoes04}. The fitness versus time curve upon periodic antimicrobial intake will be a smooth periodic function resulting from the mathematical function composition of these two empirical relationships. The main feature of this curve will be how smooth or steep  it is, which can be characterized by its rise time, i.e. the time it takes to rise from a value of $f_\mathrm{S}$ close to 0 to one close to 1. Recall that the fitness $f_\mathrm{S}$ of sensitive microorganisms ranges between 0 at very high antimicrobial concentrations and 1 without antimicrobial. In practice, we chose to define the rise time as the time taken to rise from $f_\mathrm{S}=0.1$ to $f_\mathrm{S}=0.9$.

\begin{figure}[h!]
\begin{center}
   \includegraphics[width=0.95\textwidth]{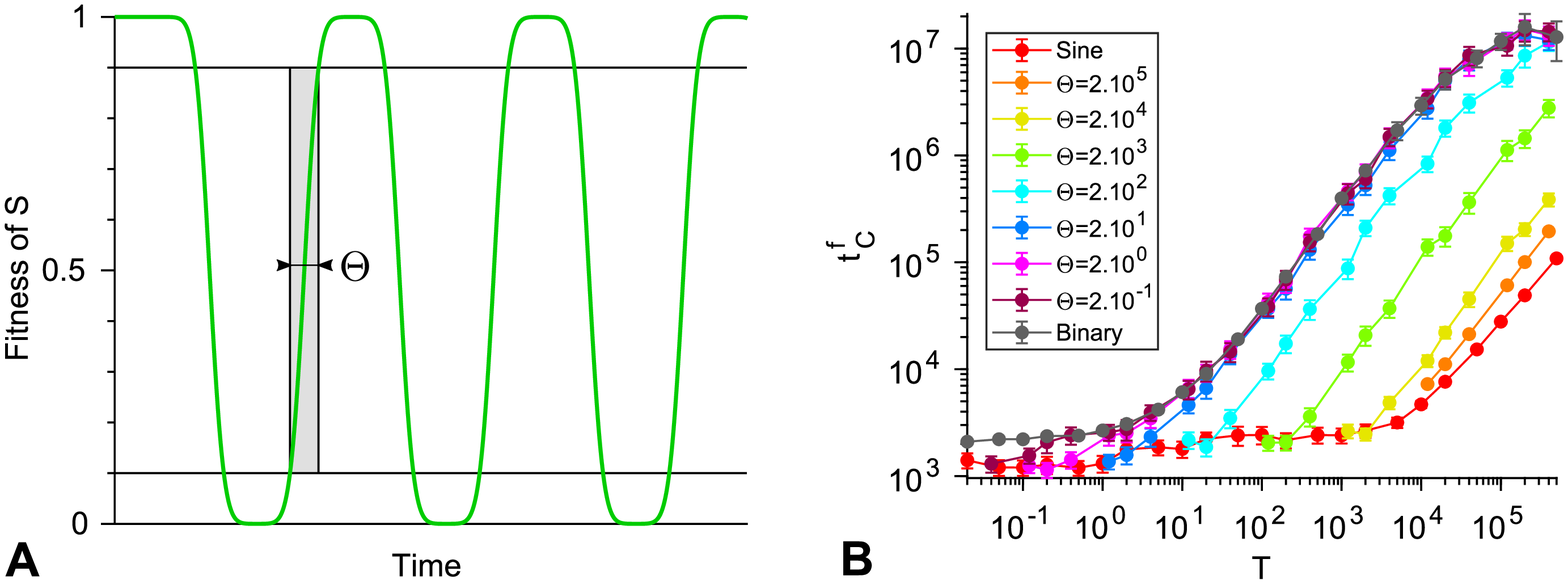}
   \quad
   \caption{\textbf{Robustness of the binary antimicrobial action model.} (A) Smooth and periodic fitness versus time relationship considered: $\Theta$ denotes the rise time. (B) Total time $t_\mathrm{C}^\mathrm{f}$ of full resistance evolution versus the period $T$ for smooth alternations with different values of $\Theta$, and for the binary model. Data points correspond to the average of simulation results (over 10 to $10^3$ replicates), and error bars (often smaller than markers) represent $95\%$ confidence intervals. Parameter values: $\mu_1=10^{-5}$, $\mu_2=10^{-3}$, $\delta=0.1$, and $N=100$.}
   \label{robustness}
\end{center}
\end{figure}

Thus motivated, we consider a smooth and periodic fitness versus time relationship $f_\mathrm{S}(t)$ (see Fig. \ref{robustness}A), and we study the impact of the rise time $\Theta$ on the evolution of antimicrobial resistance in a microbial population. In practice, our smooth function, shown in Fig.~\ref{robustness}A, is built using the error function $\mathrm{erf}(x)=(2/\sqrt{\pi})\int_0^x e^{-u^2}\, du$, such that over each period of duration $T$:
\begin{align}
f_\mathrm{S}(t)&=&1-\frac{1}{2}\left[1+\mbox{erf}\left(\frac{2}{\Theta}\left(t-nT-\frac{T}{2}\right)\right)\right]&\,\,\,\,\,\,\mbox{ if }\,\,\,nT+\frac{T}{4} \leq t < nT+\frac{3T}{4}\,,\label{erf1}\\
f_\mathrm{S}(t)&=&\frac{1}{2}\left[1+\mbox{erf}\left(\frac{2}{\Theta}\left(t-nT-T\right)\right)\right] &\,\,\,\,\,\,\mbox{ if }\,\,\,nT+\frac{3T}{4} \leq t < (n+1)T+\frac{T}{4}\,,\label{erf2}
\end{align}
where $n$ is a non-negative integer. In addition, we take $f_\mathrm{S}(t)=1$ for $0\leq t\leq T/4$, i.e. we start without antimicrobial at $t=0$, and the first decrease of fitness occurs around $t=T/2$, in order to be as close as possible to our binary approximation (see Fig.~\ref{Fig1}B). Finally, as an extremely smooth case, we consider the case of a fitness $f_\mathrm{S}$ modeled by a sine function of period $T$, with the same initial condition and phase as our function with variable smoothness.

We have performed stochastic simulations using the model described in the main text, but with the fitness versus time relationship given in Eqs.~\ref{erf1}-\ref{erf2}. Fig.~\ref{robustness} shows that for small rise times $\Theta$, the dependence on the period $T$ of the total time $t_\mathrm{C}^\mathrm{f}$ of full resistance evolution is the same as with our binary approximation, provided that the rise time is much smaller than the period, $\Theta\ll T$. Conversely, for small $\Theta$ satisfying $\Theta\geq T$, in which case our function is very smooth even though the absolute rise time is short, the behavior of $t_\mathrm{C}^\mathrm{f}$ is similar to that obtained for the sine function. For larger values of $\Theta$, namely $\Theta\gg 10$, the binary case is no longer matched when $\Theta\ll T$, and instead, a behavior intermediate between the binary case and the sine case is observed. This intermediate behavior gets closer to that observed in the sine case as $\Theta$ is increased. 

These results can be rationalized as follows. When $\Theta$ is smaller than the relevant evolutionary timescales identified in the main text ($\tau_\mathrm{R}^\mathrm{d}$, $\tau_\mathrm{R}^\mathrm{f}$ and $1/(N\mu_1)$, the shortest ones being $\tau_\mathrm{R}^\mathrm{d}$ and $\tau_\mathrm{R}^\mathrm{f}$ for $N\mu_1\ll 1$), no relevant evolutionary process process can happen during a single smooth rise or decay of the fitness. If in addition $\Theta$ is much smaller than the environmental timescale $T$, then the fitness versus time function is steep  and effectively binary. However, if $\Theta$ is not much smaller than $T$, then the function is smooth, and the binary approximation is inappropriate. Finally, if $\Theta$ is longer than the shortest relevant evolutionary timescales ($\tau_\mathrm{R}^\mathrm{d}$, $\tau_\mathrm{R}^\mathrm{f}$), then relevant evolutionary processes can happen within a single smooth rise or decay of the fitness, and the behavior is more complex. In a nutshell, our binary approximation is appropriate provided that the rise time satisfies $\Theta\ll \min (T,\tau_\mathrm{R}^\mathrm{d},\tau_\mathrm{R}^\mathrm{f})$.

\end{document}